\RequirePackage{fix-cm}
\documentclass[smallextended]{svjour3}       
\smartqed  
\usepackage{graphics,epsfig}
\usepackage{graphicx}
\usepackage{float}
\usepackage{amssymb,amstext,amsmath}
\usepackage{mathtools}
\usepackage{booktabs}
\usepackage{natbib}
\usepackage[latin1]{inputenc}
\usepackage{epstopdf}
\usepackage{xcolor}
\usepackage{verbatim}
\usepackage{physics}
\usepackage{nomencl}
\makenomenclature
\begin{document}
\newcommand{\balpha}{\boldsymbol{\alpha}}
\newcommand{\bbeta}{\boldsymbol{\beta}}
\newcommand{\bgamma}{\boldsymbol{\gamma}}
\newcommand{\bdelta}{\boldsymbol{\delta}}
\newcommand{\bepsilon}{\boldsymbol{\epsilon}}
\newcommand{\bvarepsilon}{\boldsymbol{\varepsilon}}
\newcommand{\bzeta}{\boldsymbol{\zeta}}
\newcommand{\bfoldeta}{\boldsymbol{\eta}}
\newcommand{\btheta}{\boldsymbol{\theta}}
\newcommand{\bvartheta}{\boldsymbol{\vartheta}}
\newcommand{\biota}{\boldsymbol{\iota}}
\newcommand{\bkappa}{\boldsymbol{\kappa}}
\newcommand{\blambda}{\boldsymbol{\lambda}}
\newcommand{\bmu}{\boldsymbol{\mu}}
\newcommand{\bnu}{\boldsymbol{\nu}}
\newcommand{\bxi}{\boldsymbol{\xi}}
\newcommand{\bpi}{\boldsymbol{\pi}}
\newcommand{\bvarpi}{\boldsymbol{\varpi}}
\newcommand{\brho}{\boldsymbol{\rho}}
\newcommand{\bvarrho}{\boldsymbol{\varrho}}
\newcommand{\bsigma}{\boldsymbol{\sigma}}
\newcommand{\bvarsigma}{\boldsymbol{\varsigma}}
\newcommand{\btau}{\boldsymbol{\tau}}
\newcommand{\bupsilon}{\boldsymbol{\upsilon}}
\newcommand{\bphi}{\boldsymbol{\phi}}
\newcommand{\bvarphi}{\boldsymbol{\varphi}}
\newcommand{\bchi}{\boldsymbol{\chi}}
\newcommand{\bpsi}{\boldsymbol{\psi}}
\newcommand{\bomega}{\boldsymbol{\omega}}
\newcommand{\bGamma}{\boldsymbol{\Gamma}}
\newcommand{\bDelta}{\boldsymbol{\Delta}}
\newcommand{\bTheta}{\boldsymbol{\Theta}}
\newcommand{\bLambda}{\boldsymbol{\Lambda}}
\newcommand{\bXi}{\boldsymbol{\Xi}}
\newcommand{\bPi}{\boldsymbol{\Pi}}
\newcommand{\bSigma}{\boldsymbol{\Sigma}}
\newcommand{\bUpsilon}{\boldsymbol{\Upsilon}}
\newcommand{\bPhi}{\boldsymbol{\Phi}}
\newcommand{\bPsi}{\boldsymbol{\Psi}}
\newcommand{\bOmega}{\boldsymbol{\Omega}}
\newcommand{\ldbracket}{[\![}
\newcommand{\rdbracket}{]\!]}
\def\Xint#1{\mathchoice
   {\XXint\displaystyle\textstyle{#1}}%
   {\XXint\textstyle\scriptstyle{#1}}%
   {\XXint\scriptstyle\scriptscriptstyle{#1}}%
   {\XXint\scriptscriptstyle\scriptscriptstyle{#1}}%
   \!\int}
\def\XXint#1#2#3{{\setbox0=\hbox{$#1{#2#3}{\int}$}
     \vcenter{\hbox{$#2#3$}}\kern-.5\wd0}}
\def\ddashint{\Xint=}
\def\dashint{\Xint-}

\title{Dislocation impediment by the grain boundaries in polycrystals
}


\author{Yinguang Piao         \and
        Khanh Chau Le 
}


\institute{Yinguang Piao \at
              Lehrstuhl f\"ur Mechanik - Materialtheorie, Ruhr-Universit\"at Bochum, D-44780 Bochum, Germany                        
           \and
           Khanh Chau Le$^{a,b}$ (Corresponding author) \at
$^a$\,Materials Mechanics Research Group, Ton Duc Thang University, Ho Chi Minh City, Vietnam
\\
$^b$\,Faculty of Civil Engineering, Ton Duc Thang University, Ho Chi Minh City, Vietnam\\
\email{lekhanhchau@tdtu.edu.vn}
}

\date{Received: date / Accepted: date}

\maketitle

\begin{abstract}
Thermodynamic dislocation theory incorporating dislocation impediment by the grain boundaries is developed to analyze the shear test of polycrystals. With a small set of physics based material parameters, we are able to simulate the stress-strain curves for the load and its reversal, which are consistent with the experimental curves of \citet{thuillier2009comparison}. Representative distributions of plastic slip under load and its reversal are presented, and their evolution explains the extended length of the transition stage during load reversal.
\end{abstract}

\nomenclature{$h,c,L$}{Height, width, and depth of the slab}
\nomenclature{$\gamma$}{Shear amount (as control parameter)}
\nomenclature{$e_D$}{Formation energy of one dislocation}
\nomenclature{$\gamma_D$}{Formation energy of one dislocation per unit length}
\nomenclature{$\zeta_Y$}{Yield surface tension}
\nomenclature{$\beta$}{Plastic slip}
\nomenclature{$a^2$}{The minimally possible area occupied by one dislocation}
\nomenclature{$b$}{Magnitude of Burgers' vector}
\nomenclature{$\rho$}{Total dislocation density}
\nomenclature{$\rho_i$}{Initial value of $\rho$}
\nomenclature{$\rho_g$}{Density of non-redundant dislocations}
\nomenclature{$\rho_{cr1}$}{Critical density of non-redundant dislocations impeded at grain boundaries with small misorientation angles}
\nomenclature{$\rho_{cr2}$}{Critical density of non-redundant dislocations impeded at grain boundaries with moderate misorientation angles}
\nomenclature{$\rho_r$}{Density of redundant dislocations}
\nomenclature{$\chi$}{Configurational (effective) temperature}
\nomenclature{$\chi_i$}{Initial value of $\chi$}
\nomenclature{$T$}{Kinetic-vibrational temperature}
\nomenclature{$\tau$}{Applied shear stress}
\nomenclature{$\tau_T$}{Taylor stress}
\nomenclature{$\tau_i$}{Internal stress}
\nomenclature{$\tau_b$}{Back stress}
\nomenclature{$\mu$}{Shear modulus}
\nomenclature{$\nu$}{Stress ratio}
\nomenclature{$T_P$}{Energy barrier expressed in the temperature unit}
\nomenclature{$t_0$}{Time characterizing the depinning rate}
\nomenclature{$q$}{Dimensionless plastic strain rate}
\nomenclature{$q_0$}{Dimensionless total strain rate}
\nomenclature{$\chi_0$}{Steady-state configurational temperature}
\nomenclature{$\psi_m$}{Energy of non-redundant dislocations}
\nomenclature{$\rho_{ss}$}{Steady-state dislocation density}
\nomenclature{Dot over quantities}{Time rates}
\nomenclature{Tilde over quantities}{Rescaled (dimensionless) quantities}
\printnomenclature

\section{Introduction}
Dislocations that occur during plastic deformations of polycrystalline materials can cause two types of work hardening. The first of these is isotropic work hardening due to dislocation entanglement,  determined by the kinetics of dislocation depinning and hence the rate of average plastic slip. The resultant Burgers' vector of dislocations causing this isotropic hardening over any macroscopic representative area element is zero, therefore they are called redundant (statistically stored) dislocations \citep{cottrell1964mechanical,ashby1970deformation,weertman1996dislocation,arsenlis1999crystallographic,arsenlis2004evolution}. The second of these is kinematic hardening by the accumulation of non-redundant (geometrically necessary) dislocations against obstacles in form of grain boundaries or precipitates. The resultant Burgers vector of this type of dislocations does not disappear and can be expressed in terms of the Nye's tensor   \citep{nye1953some,bilby1955types,kroner1955fundamentale,kroner1958kontinuumstheorie,mura1965continuous,berdichevsky1967dynamic,le1996model,weertman1996dislocation}. One of us has shown that for the construction of thermodynamic dislocation theory (TDT) combining both types of work hardening, the configurational entropy introduced by \citet{langer2010thermodynamic} and the density of non-redundant dislocations should be taken into account \citep{le2018athermodynamic}. Within the small strain theory he proposed the energy and dissipation of dislocated crystals and derived the governing equations of TDT exhibiting indeed both types of work hardening. Of the various dislocation based plasticity theories, we mention here only those in  \citep{ortiz1999nonconvex,groma2003spatial,berdichevsky2006continuum,berdichevsky2006thermodynamics,acharya2010new,anand2015stored,levitas2015thermodynamically,wulfinghoff2015gradient,hochrainer2016thermodynamically,chowdhury2016fluctuation,berdichevsky2019beyond,po2019continuum,lieou2020thermodynamic} which are closely relevant to our thermodynamic approach. The continuum models incorporating the grain boundaries within the small strain gradient plasticity have been proposed in \citep{gurtin2008theory,fleck2009mathematical,ekh2011influence,wulfinghoff2013gradient,voyiadjis2014theory,gottschalk2016computational}. The extension of these models to finite strains has been considered in \citep{mcbride2016computational,alipour2019grain}.

The application of the above TDT for non-uniform plastic deformation to predicting the Bauschinger effect in a single crystal (one grain) has been initiated in \citep{le2018cthermodynamic}. In that paper the grain boundaries are modelled as hard obstacles which do not allow dislocations to reach them. Due to these boundary conditions non-redundant screw dislocations that appear after being depinned will move under the applied shear stress in the opposite directions and pile up against the grain boundaries leading to the kinematic hardening. The presence of the positive back stress during the load reversal reduces the magnitude of shear stress required to pull non-redundant dislocations back to the center of the specimen. There, the non-redundant dislocations of opposite signs meet and annihilate each other leading to the Bauschinger effect. However, as observed in the experiments conducted by \citet{kondo2016direct}, low-angle grain boundaries do not always block dislocations, and when the dislocation density achieves a critical value, the non-redundant dislocations may reach or traverse the grain boundaries (cf. \citep{navarro1988alternative,lee1989prediction,koning2002modelling,britton2009nanoindentation,bayerschen2016review}). There are two possible scenarios after achieving this critical value of dislocation density: (i) Non-redundant screw dislocations on both sides of the grain boundary are of opposite signs, so when reaching the grain boundary they annihilate each other and becoming redundant dislocations with the consequence that the surface dislocation density vanishes producing no jump in the plastic slip, (ii) Non-redundant screw dislocations on both sides are of the same signs, so their densities must be equal after traversing the grain boundary. The aim of this paper is to derive the new boundary condition at the grain boundary from the variational equation within a small strain theory that modifies the condition formulated in \citep{le2018cthermodynamic}. We show that the first scenario happens in our problem that affects the stress-strain curves as well as the plastic slip distribution during loading and load reversal. We also compare our theoretical stress-strain curves with those measured in \citep{thuillier2009comparison}.

The paper is structured as follows. After this brief Introduction, we describe in Section~\ref{Section Theory} the dislocation impediment by the grain boundaries, discuss the main concept of TDT and derive its governing equations and boundary conditions for polycrystals including the condition at the grain boundaries. Section 3 develops the numerical integration of the governing equations and investigates the influence of grain boundaries on the distribution of plastic slip and the corresponding hardening behavior. Section~\ref{Section Comparison} compares the results of simulation with experiments. We conclude in Section~\ref{Section Conclusion} with a brief summary and future research directions. 

\section{Dislocation impediment by the grain boundaries and thermodynamic dislocation theory} \label{Section Theory}
\subsection{Dislocation impediment by the grain boundaries}
\begin{figure}[htb]
	\centering
	\includegraphics[width=0.6 \textwidth]{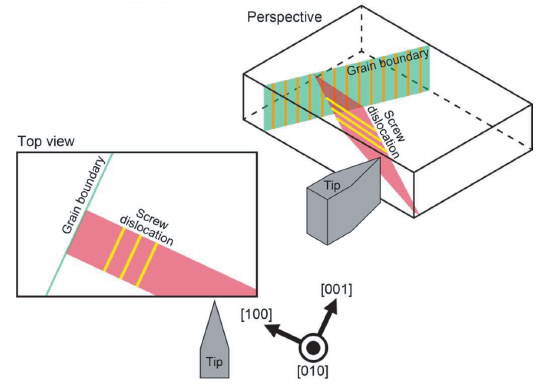}
	\caption{(Color online) Schematic illustration of the geometric arrangement of the specimen, the grain boundary, and screw dislocation. \citep{kondo2016direct}}
	\label{fig: Experiment Illustration1}
\end{figure}
\begin{figure}[htb]
	\centering
	\includegraphics[width=0.8 \textwidth]{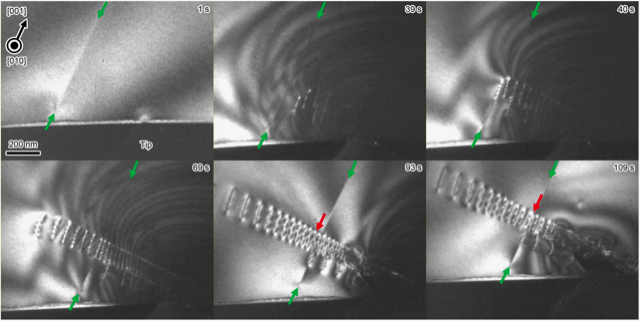}
	\caption{(Color online) Sequential transmission electron microscopy images captured from the movie of the nanoindentation experiment for the low-angle tilt grain boundary. \citep{kondo2016direct} }
	\label{fig: Experiment Illustration2}
\end{figure}

\citet{kondo2016direct} directly observed how the motion of individual dislocations is impeded at a well-defined grain boundary in strontium titanate, via in situ nanoindentation experiments. Fig.~\ref{fig: Experiment Illustration1} shows the geometric setup of the sample including the (100) low-angle grain boundary consisting of the periodic array of edge dislocations. The rotation angle of the adjacent crystal is $1.2^{\circ}$ around the $[100]$-axis. When the indenter tip is inserted into the sample, screw dislocations are emitted from the indenting point and propagate towards the grain boundary along the direction $[0\bar{1}1]$ on the $(011)$ slip plane. Fig.~\ref{fig: Experiment Illustration2} shows the dynamic process of dislocation movement during the indentation experiment in which the nucleated screw dislocations move on the slip plane, approaching and traversing the grain boundary. It was found that when the dislocations hit the grain boundary, they were slightly hindered in the core area of the grain boundary.  
Another observation in their work shows that the high-angle grain boundary, where the rotation angle of the adjacent grains is $36.9^{\circ}$, behaves as a stronger barrier such that no dislocations pass through the grain boundary before the specimen edge was fractured due to the stress at the indenting point. 

\subsection{Kinetics and thermodynamics of dislocations}
We start with the formula for the average plastic slip rate $\dot{\bar{\beta}}$ given by 
\begin{equation*}
\dot{\bar{\beta}}=\rho_r b v,
\end{equation*}
with $\rho_r$ being the density of redundant dislocation, $b$ the magnitude of the Burgers' vector, and $v$ the mean velocity of dislocations. This formula is the extension of the well-known Orowan's equation $\dot{\beta}=\rho bv$ to non-uniform plastic deformation and reduces to it in the uniform case when $\rho_r$ coincides with the total density of dislocations $\rho$. Note that the averaging is taken over one grain \citep{le2020averaging}, and that the contribution of the non-redundant dislocations to this average plastic slip rate is assumed to be negligibly small. Provided the time dislocations spend in the pinned state, $t_P$, is much longer than the time they move from one to another pinning site, the mean velocity of dislocations $v$ is given by the mean distance between them $l=1/\sqrt{\rho_r}$ multiplied by the depinning rate $1/t_P=f_P(T, \tau_i,\rho_r)/t_0$. Here, $T$ is the ordinary kinetic-vibrational temperature, $\tau_i$ the internal stress required for dislocations to overcome the resistance due to pinning (the name for this quantity was given by \citet{cottrell1953dislocations}), while $t_0$ a microscopic time of the order of the inverse Debye frequency. The activation term is given by the double exponential formula
\begin{equation}\label{activation}
f_P(T,\tau_i,\rho_r)= \exp \Bigl\{ \Bigl[-\frac{T_P}{T}\exp \Bigl[ -\frac{\tau_i}{\tau_T(\rho_r)} \Bigr] \Bigr]\Bigr\} .
\end{equation}
In the presence of the internal stress $\tau_i$ the energy barrier $e_P = k_B T_P$ is reduced by the stress-dependent factor $e^{-\tau_i/\tau_T(\rho_r)}$, where $\tau_T (\rho_r) = \mu_T b \sqrt{\rho_r}$ is the Taylor stress, while $\mu_T$ is proportional to the shear modulus $\mu$. $k_B$ and $T_P$ are the Boltzmann factor and the activation temperature, respectively. As the internal stress $\tau_i$ depends on the applied shear stress $\tau$ as will be seen later in Eq.~\eqref{TDT Balance of microforce}, the average plastic slip rate also depends on $\tau$. For the crystal loaded in both directions such that the shear rate $\dot{\gamma}$ may change its sign, the dimensionless average plastic slip rate reads
\begin{equation} \label{TDT Dimensionless strain rate}
q(T,\tau_i, \rho_r)=\dot{\bar{\beta}} t_0=b\sqrt{\rho_r} [f_P(T,\tau_i,\rho_r)-f_P(T,-\tau_i,\rho_r)].
\end{equation}
Antisymmetry with respect to the internal stress $\tau_i$ is required in \eqref{TDT Dimensionless strain rate} for dealing with the reversal process, whereby reflection symmetry must be maintained and the plastic power $\tau_i q$ must not be negative (see \citep{langer2017thermodynamic,le2019intro}). We will see in the later Section the essential role that Eq.~\eqref{TDT Dimensionless strain rate} plays in predicting the asymmetry between loads in opposite directions at different strain levels. Assuming that the system is driven at a constant shear rate $\dot{\gamma}=q_0/t_0$, we write down the following equation for the rate of change of the internal stress \citep{le2018athermodynamic}
\begin{equation}\label{Governing tau_Y}
\frac{\partial \tau_i}{\partial \gamma}= \mu \Bigl[1-\frac{q(T,\tau_i, \rho_r)}{q_0}\Bigr].
\end{equation}

The evolution equation of the effective (disorder) temperature follows from the first law of thermodynamics applied to the configurational subsystem of dislocations \citep{langer2010thermodynamic,langer2017thermodynamic}. The basic idea of the first law for the theory of effective temperature is as follows. Assuming that the system has neither an external heat source inside the body, nor a heat flow through the material surface supplying external energy to the body, then the only input power is the rate of external work, and it is balanced at the rate of total internal energy, which is the sum of the internal energy rates of two subsystems. The mechanical power supplied is partially stored in the form of dislocations, and the remainder is converted into heat flow in two subsystems. Let $S_R$ and $S_C$ be the entropy of the kinetic-vibrational and configurational subsystem, respectively. Part of this remaining power is the heat flux $\theta \dot{S}_R$, which flows from the configuration subsystem into the kinetic-vibrational  subsystem during the irreversible rearrangement of the atoms.  Since the fast kinetic-vibrational degrees of freedom are combined with the external environment to serve as a single thermal bath at a fixed ordinary temperature $T$, the loss of thermal energy through the material surface is included in this heat flux. The remaining part is the rate of configurational heat $\chi \dot{S}_C$ \citep{langer2017thermodynamic}, and therefore it is the difference between the applied plastic power and the total rate of energy input to dislocations and kinetic-vibrational heat. The configurational heat is reformulated in relation to the rate of effective temperature by introducing an effective specific heat $c_{eff}$. The evolution equation of the effective temperature is (see the detailed derivation in \citep{langer2017thermodynamic,le2019intro})
\begin{equation}\label{Governing chi}
\frac{\partial \chi}{\partial \gamma}=K_{\chi}\frac{\tau_i  e_D  q}{\mu  q_0}\Bigl[1-\frac{\chi}{\chi_0}\Bigr].
\end{equation}
Here $e_D$ is the formation energy of one dislocation, while $\chi_0$ is the steady-state value of $\chi$ for strain rates which are significantly smaller than the inverse atomic relaxation time, i.e. much smaller than $t_0^{-1}$. The dimensionless factor $K_{\chi}$ is proportional to the inverse of $c_{eff}$.

The second law of thermodynamics, from which the evolution equation of dislocation density is derived, requires that the sum of the entropy rate of two subsystems is non-negative. By evaluating the entropy of the configurational subsystem from the first law and replacing it with the second law, the reformulation of the second law is obtained and a guideline for the form of the dislocation density rate is provided. With careful derivation (see \citep{langer2017thermodynamic} for details) we write the evolution equation for the total dislocation density $\rho=\rho_r+\rho_g$ as
\begin{equation}\label{Governing rho}
\frac{\partial \rho}{\partial \gamma}=K_{\rho} \frac{\tau_i  q}{a^2 \mu \nu^2 q_0} \Bigl(1-\frac{\rho}{\rho_{ss}(\chi)}\Bigr),
\end{equation}
where $\rho_g$ is the density of non-redundant dislocations. The coefficient $K_{\rho}$, assumed to be independent of both strain rate and temperature, is an energy conversion factor from the applied mechanical work into dislocations. $\rho_{ss}(\chi)=(1/a^2)e^{-e_D/\chi}$ is the steady-state dislocation density determined by minimization of free energy of configurational subsystem, with  $a^2$ denoting the minimally possible area occupied by one dislocation. Function $\nu(T,\rho_r, q)$ arises from solving Eq.~\eqref{TDT Dimensionless strain rate} for the internal stress as a function of the dimensionless strain rate,
\begin{equation*}
\nu(T,\rho_r, q)= \ln\Bigl(\frac{T_P}{T}\Bigr)-\ln\Bigl[\ln\Bigl(\frac{b\sqrt{\rho_r}}{q}\Bigr)\Bigr]
\end{equation*}
and has the meaning of the stress ratio $\tau_i/\tau_T$.

The equation for the plastic slip $\beta$ reads (cf. \citep{le2018athermodynamic})
\begin{equation} \label{TDT Balance of microforce}
\tau-\tau_b-\tau_i=0, 
\end{equation}
where $\tau$ is the applied shear stress, $\tau_b$ is the back stress originating in gradient plasticity (see, e.g., \citep{le1996model,gurtin2008theory}) and describing the interaction between non-redundant dislocations, and the internal stress $\tau_i$ is attributed to the density of redundant dislocations \citep{cottrell1953dislocations,cottrell1964mechanical}. Physically, Eq.~\eqref{TDT Balance of microforce} is interpreted as the balance of microforces acting on dislocations. It should be noted, however, that this equation differs substantially from the similar equations of balance of microforces proposed in the above cited papers \citep{le1996model,gurtin2008theory}, in that the internal stress itself is a dynamic variable that evolves along with the dislocation density according to Eq.~\eqref{Governing tau_Y}. A derivation of Eq.~\eqref{TDT Balance of microforce} is given in the next Section and with further details in the Appendix.
 
\begin{figure}[htb]
	\centering
	\includegraphics[width=0.5 \textwidth]{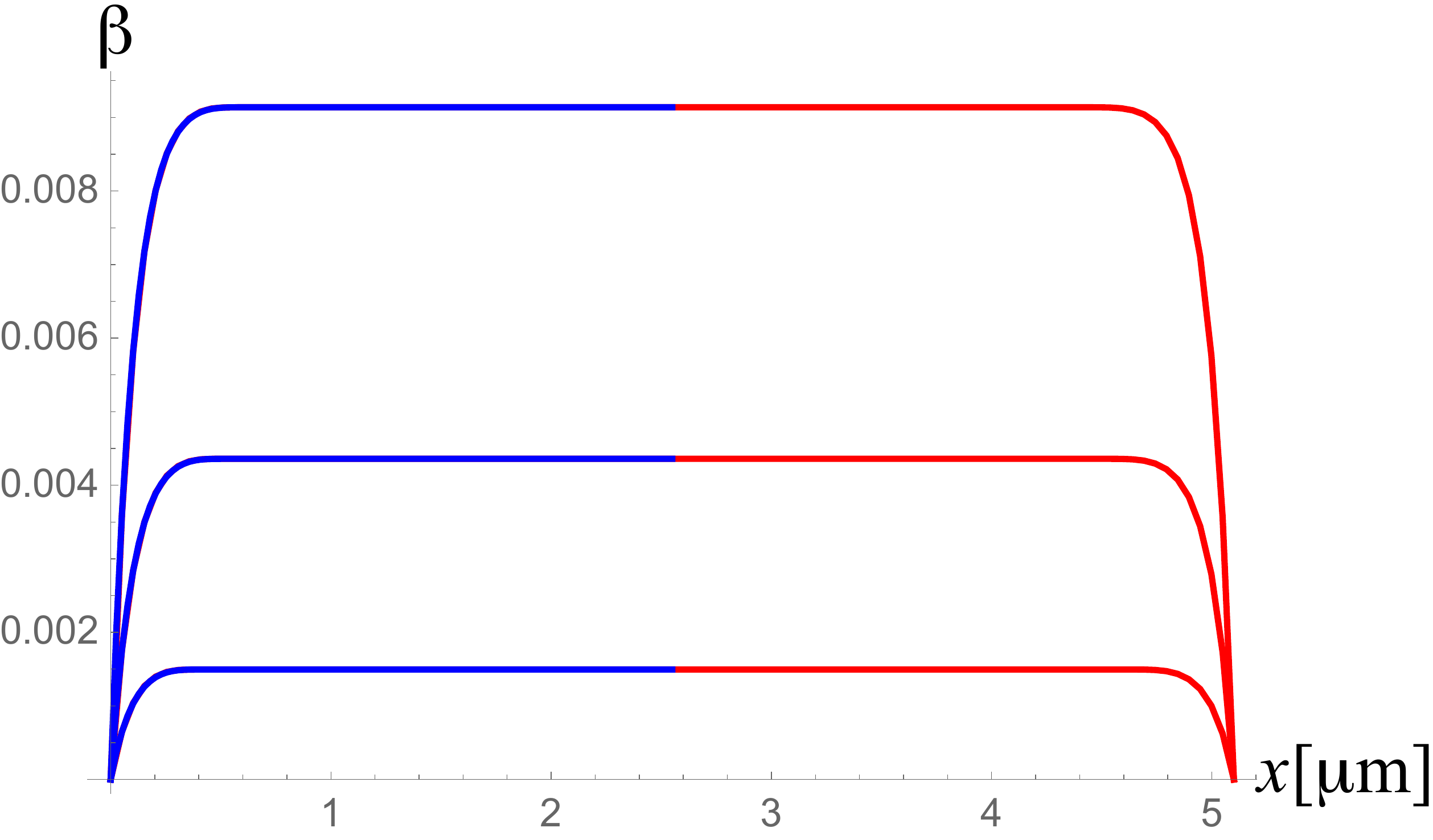}
	\caption{(Color online) Comparison of the plastic slip obtained from Dirichlet-Dirichlet boundary condition in the interval $x\in{(0, c)}$ (blue and red) and from Dirichlet-Neumann boundary condition in the interval $x\in{(0, c/2)}$ (blue).}
	\label{fig:Dirichlet-Neumann}
\end{figure}

Evolution equations for the internal stress, effective temperature, dislocation density and plastic slip form a system of coupled partial differential equations. All variables $\tau_i(x,\gamma)$, $\chi(x,\gamma)$, $\rho(x,\gamma)$ and $\beta(x,\gamma)$ have initial conditions, while $\beta(x,\gamma)$ is subject to certain boundary conditions. 
\citet{le2018cthermodynamic} investigated the Bauschinger effect of a single crystal deforming in anti-plane shear. Since the prescribed displacement does not allow non-redundant dislocations to reach the boundary, the homogeneous Dirichlet condition of plastic slip, $\beta=0$, is applied at the outer boundaries (see Figure 1 in \citep{le2018cthermodynamic}). In their work, the plastic slip at a given shear strain decreases rapidly to zero at two boundary layers and is constant in the middle. This boundary condition leads to dislocation accumulation zones at two boundary layers and a dislocation-free zone between these two layers. \citet{piao2020thermodynamic} added an imaginary boundary located at $x=c/2$ and posed the boundary conditions in the interval $x\in{(0, c/2)}$ as
\begin{equation}\label{Dirichlet-Neumann BC}
\beta(0,\gamma)=0, \quad 
\beta_{,x}(c/2,\gamma)=0.
\end{equation}
If we compare the distribution of plastic slip obtained from Eq.~\eqref{Dirichlet-Neumann BC} in the interval $(0, c/2)$ (blue curves) with that obtained from the homogeneous Dirichlet-Dirichlet boundary conditions in the interval $(0, c)$ (blue and red curves), we see from Fig.~\ref{fig:Dirichlet-Neumann} that they coincide in $(0, c/2)$. It turns out that a boundary subjected to the homogeneous Neumann boundary condition can be considered a free boundary, through which dislocations can pass freely. Similarly, a non-homogeneous Neumann boundary condition indicates the accumulation of non-redundant dislocations at the grain boundary. We will derive the grain boundary conditions in the following sub-Section.

\subsection{Conditions at the grain boundary and role of dislocation impediment}

\begin{figure}[htb]
	\includegraphics[width=0.8 \textwidth]{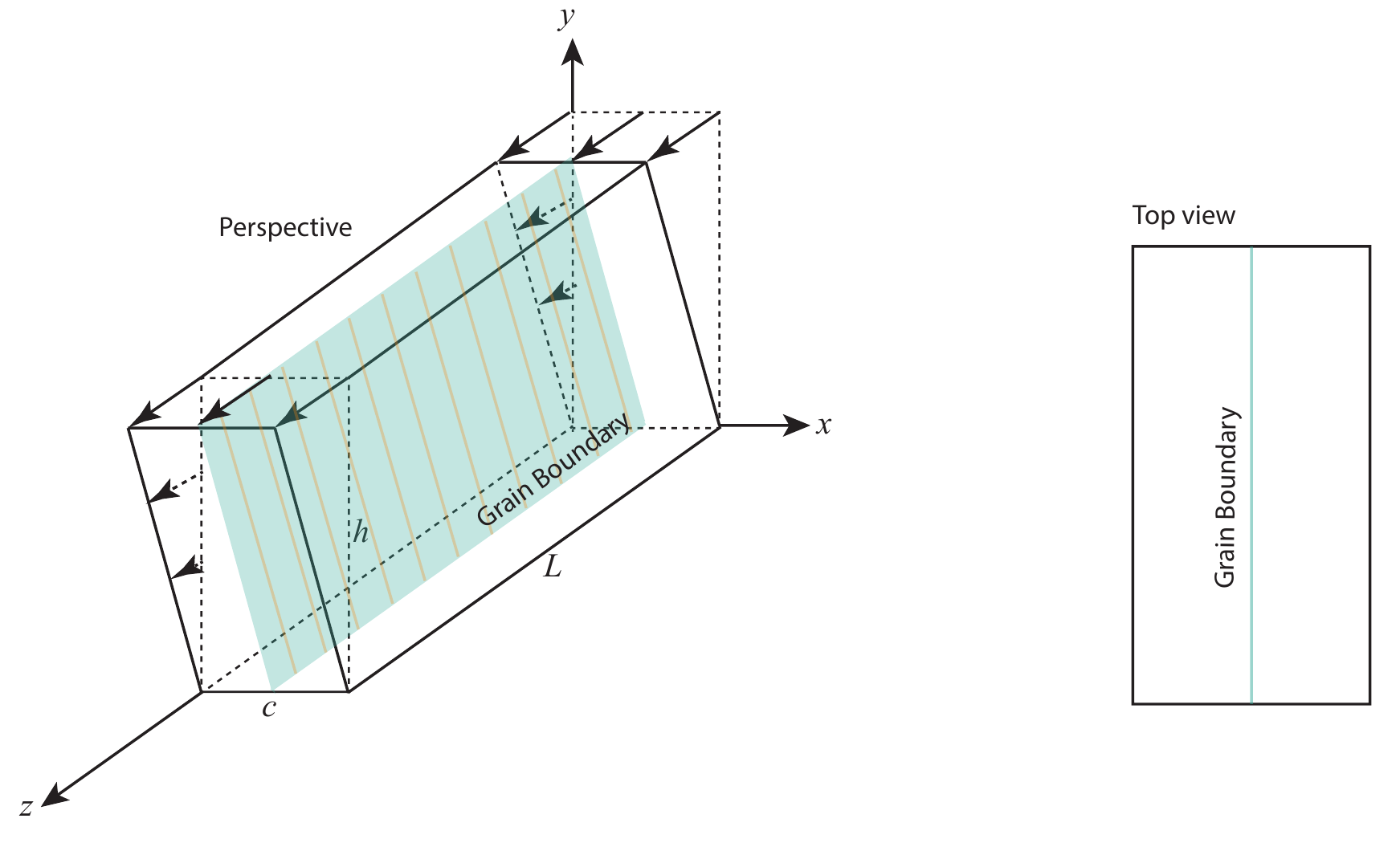} 
	\caption{(Color online) Perspective and top view of a bicrystal under anti-plane constrained shear with a grain boundary in the middle.}
	\label{fig: Antiplane Model}
\end{figure}

Motivated by the experimental observation of the dislocation impediment by a low-angle grain boundary \citep{kondo2016direct}, we investigate in this sub-Section a bicrystal with one grain boundary undergoing anti-plane constrained shear for which the displacement vector has only one non-zero component 
\begin{displaymath}
u_x=u_y=0,\quad u_z\ne 0.
\end{displaymath}
Let the cross section of this bicrystals, perpendicular to $z$-axis, be a rectangle with the width $c$ and height $h$ (see Fig.~\ref{fig: Antiplane Model}). The width is assumed much smaller than the height, while the latter is much smaller than the depth: $c\ll h\ll L$. Due to this assumption the end effects at $y=0$ and $y=h$ can be neglected, and the independent variables are reduced to $x$ and $t$. The plane of grain boundary is parallel to $(y,z)$-plane and located at $x=c/2$. We use the same material model as well as parameters for TDT presented in \citep{le2018cthermodynamic}. However, instead of the Dirichlet boundary conditions at the external  boundaries, we pose the mixed boundary conditions where the out-of-plane displacement at the upper, lower and left boundaries is prescribed as $u_z=\gamma(t) y$, with $\gamma(t)$ being the given time-dependent shear strain, while the right boundary is set as a free surface. The screw dislocations whose lines are parallel to $z$-axis may appear during this shear deformation. We assume that the plastic slip depends only on $x$ and $t$: $\beta=\beta(x,t)$. Because of the prescribed displacement at the boundary $x=0$, dislocations cannot reach that boundary, so we put $\beta(0,t)=0$. We want to study how this grain boundary influences the distribution of plastic slip, as well as the hardening behavior.

We turn now to the derivation of the governing equations and boundary conditions of TDT from the variational equation. If the dislocations cannot reach the grain boundary due to the above mentioned impediment, the plastic slip must vanish at $x=c/2$. In this case the non-redundant dislocations of opposite signs pile up on both sides of the grain boundary. When dislocation densities become sufficiently large, the grain boundary can no longer block dislocations. Then the non-redundant dislocations of opposite signs meet and cancel each other at the grain boundary, becoming redundant dislocations so that the surface dislocation density disappears with the consequence that the jump in $\beta$ at $x=c/2$ remains zero all the time. The plastic slip $\beta(c/2,t)$ can however change over time as will be seen  later in Figure~\ref{Plastic Slips}. Within TDT the equations and boundary condition governing the evolution of the plastic slip $\beta(x,t)$, the total dislocation density $\rho(x,t)=\rho_r+\rho_g$, and the disorder temperature $\chi(x,t)$ can be derived from the variational equation. To do this we have to propose the energy and dissipation potential. Since there is no change in the surface dislocation density and therefore the surface energy can be neglected, we let the energy functional, normalized by $hL$, be exactly the same as in \citep{le2018cthermodynamic} 
\begin{equation}\label{Free_energy_density}
I[\beta,\rho_r,\bar{\chi}]=\int_0^c\Bigl[\frac{1}{2}\mu (\gamma -\beta)^2+\gamma_D\rho_r+\psi_m(\rho_g)-\bar{\chi }(-\rho \ln (a^2\rho )+\rho )\Bigr]\dd{x},
\end{equation}
with $\gamma_D=e_D/L$ being the energy of one dislocation per unit length and $\rho_g=|\beta_{,x}|/b$. The first term on the right hand side of Eq.~\eqref{Free_energy_density} is the energy due to the elastic strain, with $\mu$ being the shear modulus (for simplicity of the analysis, the crystal is assumed to be elastically isotropic). The second term is the self-energy of redundant dislocations. The third term $\psi_m$ is the energy of non-redundant dislocations, while the last one is the configurational heat \citep{langer2017thermodynamic}. The neglect of this last term containing $\bar{\chi}$ in Eq.~(8) and consequently the absence of Eq.~(4) for it would contradict the second law for plastic flow: The configurational entropy of the subsystem of dislocations must increase \citep{le2020two,langer2020scaling}. Note that, due to the planar distribution of dislocations, $S_C=-\rho \ln (a^2\rho )+\rho$ is the configurational entropy per unit area, so its dual quantity $\bar{\chi}$ must be interpreted as the ``two-dimensional'' configurational temperature. To get the temperature having the unit of energy, we may define $\chi=L\bar{\chi}$. \citet{berdichevsky2017acontinuum} has found $\psi_m$ for the locally periodic arrangement of non-redundant screw dislocations in a bar under torsion that agrees with the numerical simulation provided by \citet{weinberger2011structure}. However, as shown in \citep{le2018non}, Berdichevsky's energy must be extrapolated to the extremely small or large dislocation densities to guarantee the existence of solution within TDT. The extrapolated energy density of non-redundant dislocations reads \citep{le2018non}
\begin{equation*}
\psi_m(\rho_g)=\mu b^2 \rho_g \Bigl( \phi_*+\frac{1}{4\pi }\ln \frac{1}{k_0+b^2\rho_g}\Bigr) +\frac{1}{8\pi}\mu k_1(b^2\rho_g)^2,
\end{equation*}
with $\phi_* $ being a parameter depending on the periodic dislocation structure (for the hexagonal periodic dislocation structure $\phi_* =-0.105$ \citep{berdichevsky2017acontinuum}), $k_0$ a small constant correcting the behavior of the derivative of energy at $\rho_g = 0$, and $k_1$ another constant correcting the behavior of the energy at large densities of the non-redundant dislocations.

In addition to the energy, we introduce the bulk dissipation potential in accordance with \citep{le2018athermodynamic}
\begin{equation} \label{TDT Dissipation function}
D_b(\dot{\beta}, \dot{\rho}, \dot{\bar{\chi}})=\tau_i \dot{\beta}+\frac{1}{2}d_{\rho}\dot{\rho}^2+\frac{1}{2}d_{\chi}\dot{\bar{\chi}}^2,
\end{equation}
where the coefficients $d_{\rho}$ and $d_{\chi}$ must be chosen so that the governing equations of TDT reduce to those in \citep{langer2010thermodynamic} for uniform deformations. However, in the presence of grain boundary there is an additional surface dissipation which takes into account the above mentioned dislocation impediment 
\begin{equation*}
D_s(\dot{\beta}(c/2))=\zeta_Y |\dot{\beta}(c/2,t)|/b,
\end{equation*}
where $\zeta_Y$ plays the role of the yield surface tension. Note that a similar surface dissipation term has been proposed within the gradient plasticity by \citet{poh2013homogenization}. We derive the governing equations from the variational equation
\begin{equation} \label{Sedov_equation}
\delta I+\int_{0}^c \Bigl( \frac{\partial D_b}{\partial \dot{\beta}}\delta \beta +\frac{\partial D_b}{\partial \dot{\rho}}\delta \rho + \frac{\partial D_b}{\partial \dot{\bar{\chi}}}\delta \bar{\chi} \Bigr) \dd{x} +\pdv{D_s}{\dot{\beta}}\delta \beta(c/2,t)=0.
\end{equation}
In this context the similar theory proposed by \citet{aifantis2005role} should be mentioned, in which the surface energy term depending on $\beta(c/2,t)$ is introduced in \eqref{Free_energy_density}. From the thermodynamic point of view such a term makes no sense because $\beta(c/2,t)$ does not represent the surface dislocation density, is also history-dependent and cannot be the state variable. On the contrary, the introduction of the surface dissipation potential as a function of $\dot{\beta}(c/2,t)$ agrees with the second law of thermodynamics \citep{lemaitre2000mechanics} and leads, as we will see, to the reasonable condition at the grain boundary. The variational formulations involving such dissipation potentials similar to Eq. \eqref{Sedov_equation} are considered in \citep{ganghoffer2007differential,berdichevsky2009variational}. 

It is easy to show that the standard calculus of variations, with the appropriately chosen $d_\rho$ and $d_\chi$, yields the system of equations \eqref{Governing chi}, \eqref{Governing rho}, and \eqref{TDT Balance of microforce} (see \citep{le2018athermodynamic} and the Appendix).
The variation of $\beta $ with the subsequent integration by part using $\beta(0,t)=0$ and the account of \eqref{Governing rho} reduces \eqref{Sedov_equation} to
\begin{multline}
\label{reduced_variation}
\Bigl( \pdv{\psi_m}{\rho_g}\text{sign}\beta_{,x}\Bigr|_{c/2-0} -\pdv{\psi_m}{\rho_g}\text{sign}\beta_{,x}\Bigr|_{c/2+0} -\gamma_D\text{sign}\beta_{,x}\Bigr|_{c/2-0} + \gamma_D\text{sign}\beta_{,x}\Bigr|_{c/2+0}
\\
+\zeta_Y \text{sign}\dot{\beta}(c/2,t)\Bigr) \delta \beta(c/2,t)
+\Bigl(\pdv{\psi_m}{\rho_g}-\gamma_D\Bigr) \text{sign}\beta_{,x}\Bigr|_{c} \delta \beta (c,t)=0.
\end{multline}
As the dislocation density can have different values on either side of the grain boundary, the vertical line followed by $c/2\pm 0$ indicates the limits of the preceding expression as $x$ approaches $c/2$ from the right and left, respectively. Since $\delta \beta (c,t)$ can be chosen independently and arbitrarily, we get the condition at $x=c$
\begin{equation*}
\pdv{\psi_m}{\rho_g}\Bigr|_{x=c} =\gamma_D.
\end{equation*}
If $\gamma_D$ is of the order $\mu b^2$, this equation implies that $\rho_g(c)$ equals a negligibly small number. Neglecting this number, we obtain Neumann boundary condition at $x=c$
\begin{equation*}
\beta_{,x}(c,t)=0.
\end{equation*}
Then Eq.~\eqref{reduced_variation}, with the last term being removed, implies that
\begin{equation}
\label{grain_bc}
\pdv{\psi_m}{\rho_g}\Bigr|_{c/2+0}+\pdv{\psi_m}{\rho_g}\Bigr|_{c/2-0}=\zeta_Y+2\gamma_D,
\end{equation}
provided $\text{sign}\beta_{,x}|_{c/2+0}=1$, $\text{sign}\beta_{,x}|_{c/2-0}=-1$ and $\text{sign}\dot{\beta}(c/2,t)=1$. Since the sign of screw non-redundant dislocations matches the sign of $\beta_{,x}$ and the plastic slip increases during the loading, these conditions correspond to scenario (i) discussed in the Introduction, which is taken for granted. If the left-hand side is smaller than $\zeta_Y+2\gamma_D$, then $\dot{\beta}(c/2,t)=0$. It is straightforward to extend this derivation to higher dimensions and multiple grain boundaries in polycrystals.

Condition \eqref{grain_bc} simplifies in the case $\pdv{\psi_m}{\rho_g}|_{c/2+0}=\pdv{\psi_m}{\rho_g}|_{c/2-0}$. In this case it can be solved with respect to $\rho_g$ yielding
\begin{equation} \label{Constraint}
\begin{cases}
\dot{\beta}(c/2,t)=0 & \text{as long as $\rho_g<\rho_{cr1}$},\\
|\beta_{,x}(c/2\pm 0,t)|=b\rho_{cr1} & \text{otherwise},
\end{cases}
\end{equation}
where $\rho_{cr1}$ is the root of the equation $\pdv{\psi_m}{\rho_g}=\zeta_Y/2+\gamma_D$ interpreted as the critical density of non-redundant dislocations impeded at the grain boundary. \citet{jiang2020stress} showed the average distribution of densities of redundant and non-redundant dislocations in relation to the distance from the grain boundary when the plastic strain is equal to 0.002. The density of redundant dislocations at the grain boundary is in the order of 16, while the density of non-redundant dislocations is much lower and its order is about 13. Based on this result, we first set $\rho_{cr1}=9.2 \times 10^{13}$m$^{-2}$ for numerical examples, and later we discuss how to determine the correct value by comparison with experiments. 

\section{Numerical integration and influence of the grain boundaries}
\subsection{Numerical integration}
For the purpose of numerical integration of the system of equations \eqref{Governing tau_Y}-\eqref{TDT Balance of microforce} it is convenient to introduce the dimensionless variables and quantities
\begin{gather*}
\tilde{x}=x/b, \quad \tilde{\rho}=a^2 \rho, \quad \tilde{\rho}_r=a^2 \rho_r,\quad \tilde{\chi}=\chi/e_D,  
\\
\tilde{\tau}=\tau/\mu, \quad \tilde{\tau}_i= \tau_i/\mu,\quad \tilde{\tau}_b=\tau_b/\mu, \quad \theta=T/T_P.
\end{gather*}
The variable $\tilde{x}$ changes from zero to $\tilde{c}=c/b$, where the magnitude of Burgers vector is assumed to be $b=0.25 $ nm. We rewrite the dimensionless plastic slip rate as
\begin{equation*}
q(T,\tau_i, \rho_r)= \frac{b}{a}\tilde{q}(\theta,\tilde{\tau}_i, \tilde{\rho}_r),
\end{equation*}
where
\begin{equation*}
\tilde{q}(\theta,\tilde{\tau}_i, \tilde{\rho}_r)=\sqrt{\tilde{\rho}_r}[\tilde{f}_P(\theta,\tilde{\tau}_i, \tilde{\rho}_r)-\tilde{f}_P(\theta,-\tilde{\tau}_i, \tilde{\rho}_r)].
\end{equation*}
We set $\tilde{\mu}_T=(b/a)\mu_T=\mu r$ and assume that $r$ is independent of temperature and strain rate. Then
\begin{equation*}
\tilde{f}_P(\theta,\tilde{\tau}_i, \tilde{\rho}_r)= \exp \Bigl[-\frac{1}{\theta} e^{-\tilde{\tau}_i/(r\sqrt{\tilde{\rho}_r})}\Bigr].
\end{equation*}
We define $\tilde{q}_0=(a/b)q_0$ so that $q/q_0=\tilde{q}/\tilde{q}_0$. Function $\nu$ becomes
\begin{equation*}
\tilde{\nu}(\theta,\tilde{\rho}_r,\tilde{q}_0)=\ln{\Bigl(\frac{1}{\theta}\Bigr)}- \ln\Bigl[ \ln\Bigl(\frac{\sqrt{\tilde{\rho}_r}}{\tilde{q}_0}\Bigr)\Bigr].
\end{equation*}
The dimensionless steady-state quantities are 
\begin{equation*}
\tilde{\rho}_{ss}(\tilde{\chi})=e^{-1/\tilde{\chi}}, \quad \tilde{\chi}_0 = \chi_0/e_D.
\end{equation*}
Using $\tilde{q}$ instead of $q$ as the dimensionless measure of plastic strain rate means that we are effectively rescaling $t_0$ by a factor $b/a$. For purposes of this analysis, we assume that $(a/b)t_0=10^{-12}$s. In terms of the introduced dimensionless quantities the governing equations read
\begin{equation}
\begin{split}
&\frac{\partial \tilde{\tau}_i}{\partial\gamma } = \Bigl[1-\frac{\tilde{q}(\theta,\tilde{\tau}_i,\tilde{\rho }_r)}{\tilde{q}_0}\Bigr], \\
&\frac{\partial \tilde{\rho }}{\partial \gamma }= K_\rho \frac{\tilde{\tau}_i \, \tilde{q}}{\tilde{\nu}(\theta,\tilde{\rho }_r,\tilde{q}_0)^2 \, \tilde{q}_0}\Bigl[ 1-\frac{\tilde{\rho }}{\tilde{\rho }_{ss}(\tilde{\chi })} \Bigr],  \\
&\frac{\partial \tilde{\chi }}{\partial\gamma }= K_{\chi}  \frac{\tilde{\tau}_i \, \tilde{q}}{\tilde{q}_0} \Bigl[ 1-\frac{\tilde{\chi }}{\tilde{\chi }_0(\tilde{q})} \Bigr], \label{eq:governing} \\
&\ \tilde{\tau}-\tilde{\tau}_b-\tilde{\tau}_i=0.  \\
\end{split}
\end{equation}
where 
\begin{equation}
\tilde{\tau}_b=-\frac{k_1\xi^2+(2k_0k_1-1)\xi+k_1k_0^2-2k_0}{4\pi(k_0+\xi)^2}\beta_{,\tilde{x}\tilde{x}}, \quad \text{and}\quad \xi=|\beta_{,\tilde{x}}|.
\label{eq:backstress}
\end{equation}

This system of PDEs comprises four equations in which both spatial and temporal derivatives occur. In order to achieve a numerically accurate solution, the original problem is parceled out into a large number of more easily solvable ODEs which, in their turn, are discretized in $\gamma$. Let us illustrate the discretization for the case of one grain boundary situated at $\tilde{x}=\tilde{c}/2$. The interval $0<\tilde{x}<\tilde{c}$ is first decomposed into $2n$ subintervals of the length $\Delta \tilde{x}=\tilde{c}/2n$. The first and second spatial derivatives of the plastic slip $\beta (\tilde{x},\tilde{\gamma})$ can then be calculated using the finite difference approximations
\begin{equation}
\pdv{\beta}{\tilde{x}}=\frac{\beta_{i+1}-\beta_{i}}{\Delta\tilde{x}},\quad\pdv[2]{\beta}{{\tilde{x}}}=\frac{\beta_{i+1}-2\beta_i+\beta_{i-1}}{{\Delta\tilde{x}}^2},
\label{finite_difference}
\end{equation}
where $\beta_i=\beta(i \, \Delta \tilde{x},\gamma)$. In Eq.~\eqref{eq:backstress} for the dimensionless back stress $\beta_{,\tilde{x}}$ and $\beta_{,\tilde{x}\tilde{x}}$ are computed in accordance with \eqref{finite_difference}. This back stress enters equation \eqref{eq:governing}$_4$. At $\tilde{x}=\tilde{c}/2$ this equation must be replaced by the equation
\begin{equation*}
\beta(n \Delta \tilde{x},\gamma_i)=\beta(n \Delta \tilde{x},\gamma_{i-1}),
\end{equation*}
with $\gamma_i=i\, \Delta \gamma$, as long as $|\beta_{n-1}-\beta_{n}|=|\beta_{n+1}-\beta_n|< b^2  \rho_{cr1} \Delta{\tilde{x}}$, and 
\begin{displaymath}
\beta_{n-1}-\beta_{n}=\beta_{n+1}-\beta_n=b^2  \rho_{cr1} \Delta{\tilde{x}}
\end{displaymath}
otherwise. It is easy to extend this method of discretization to the case with several grain boundaries.

Altogether, this procedure leads to a system of $8n$ ordinary differential-algebraic equations (DAE), which only have first derivatives with respect to $\gamma$. In the present study, a spatial discretization of the interval $(0,\tilde{c})$ into $2n=1000$ subintervals as well as a temporal decomposition with a step size of $\Delta\gamma=10^{-6}$ is applied, whereby the latter is to be interpreted as a numerical shear increment. Finally, the usual DAE-system is solved with the internal Matlab subroutine \textit{ode15s}.

During the load reversal we need to reverse the equation for the plastic slip rate according to
\begin{equation*}
\bar{q}(\theta,\tilde{\tau}_i,\tilde{\rho}_r)=\sqrt{\tilde{\rho}_r}[\tilde{f}_P(\theta,-\tilde{\tau}_i,\tilde{\rho}_r)-\tilde{f}_P(\theta,\tilde{\tau}_i,\tilde{\rho}_r)].
\end{equation*}
The solution is obtained for any load direction by integrating the above system with the corresponding $\tilde{q}$ or $\bar{q}$. In addition to the fulfillment of equations, the continuity requirements at the transition points of the sections must be met in order to ensure the physical consistency of the solution. Therefore the calculated end values of a section are taken as initial values for the following section.

\subsection{Influence of the grain boundaries}
\begin{figure}[htp]
	\centering
	\includegraphics[width=0.5 \textwidth]{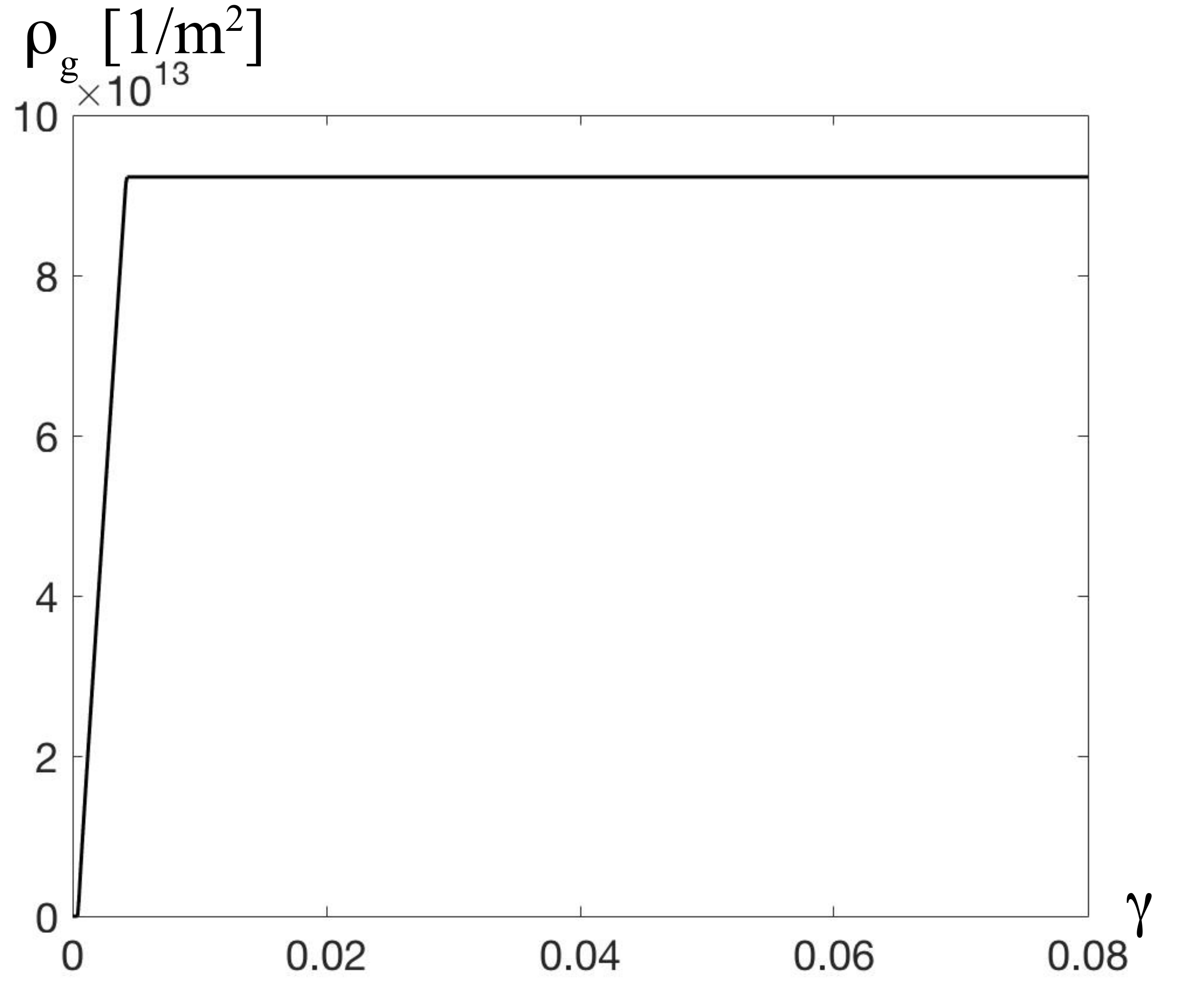}
	\caption{Evolution of non-redundant dislocation density $\rho_g$ at grain boundary with respect to shear strain $\gamma$.}
	\label{DensityMid}
\end{figure}

Low angle grain boundaries act as weak barriers to which screw dislocations in the crystal easily pile up, while non-redundant dislocation density cannot increase indefinitely. Instead, it reaches a critical value and maintains this density at the grain boundary despite the increasing strain. These features of dislocation impediment by the grain boundaries are captured by Eqs.~\eqref{Constraint}. Its first equation describes the process by which dislocations pile up at the grain boundary where the non-redundant dislocation density increases from zero to the critical value, while its second equation represents the process of dislocation traversal where dislocations of opposite signs reach the grain boundary, annihilating each other, and maintaining the critical density. This behavior is consistent with the research based on a quasi-continuum method \citep{yu2012interactions} showing that dislocation transfer does not occur until the accumulated residual defects reach a threshold. Figure~\ref{DensityMid} shows this tendency: The non-redundant dislocation density at the grain boundary increases linearly with increasing strain and remains constant after $\gamma=0.0032$. Note that the density can increase non-linearly by choosing a Neumann boundary condition whose right-hand side is a non-linear function of $\gamma$. 
\begin{figure}[htp]
	\begin{tabular}{cc}
		\includegraphics[width=0.47 \textwidth]{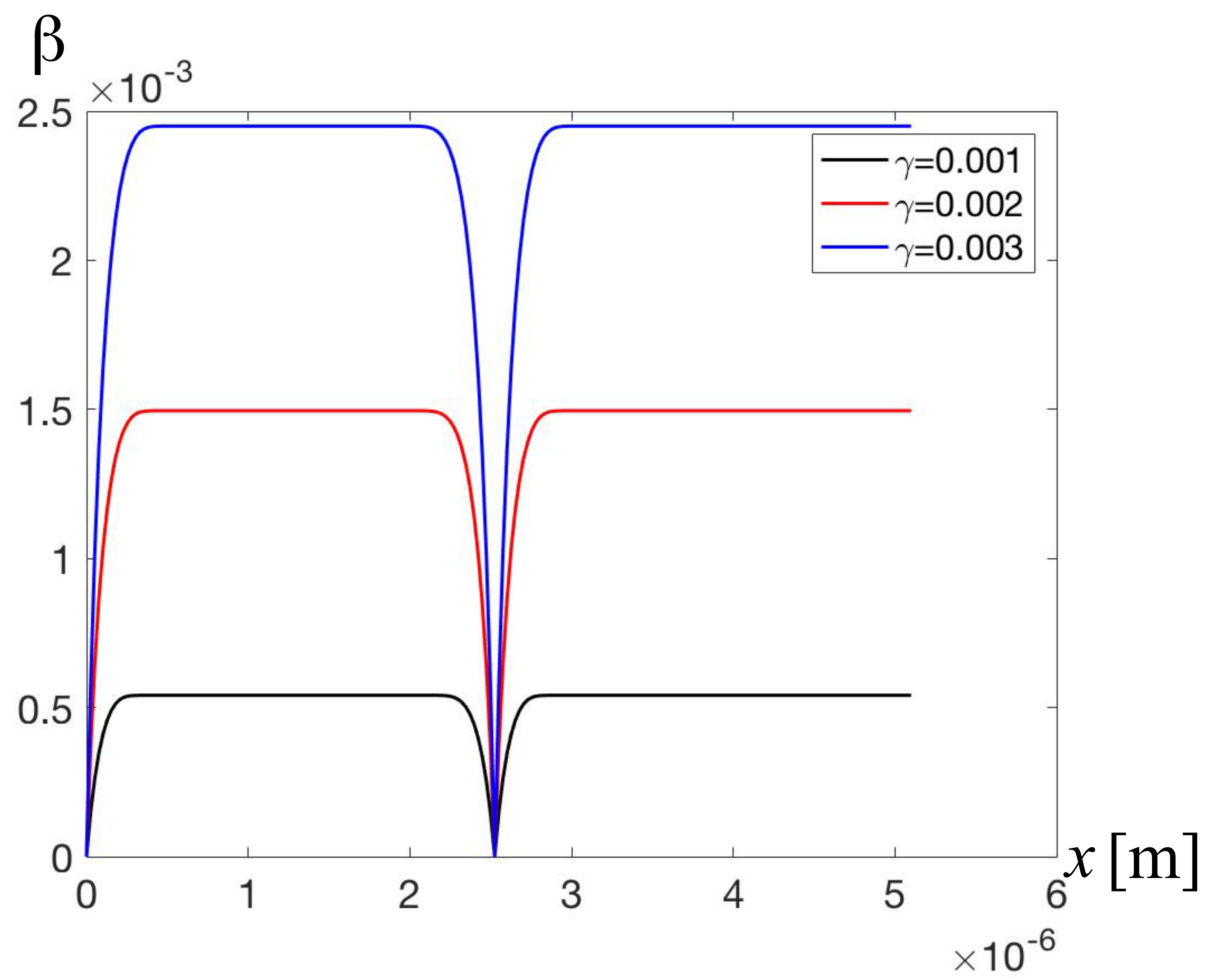} &  
		\includegraphics[width=0.48 \textwidth]{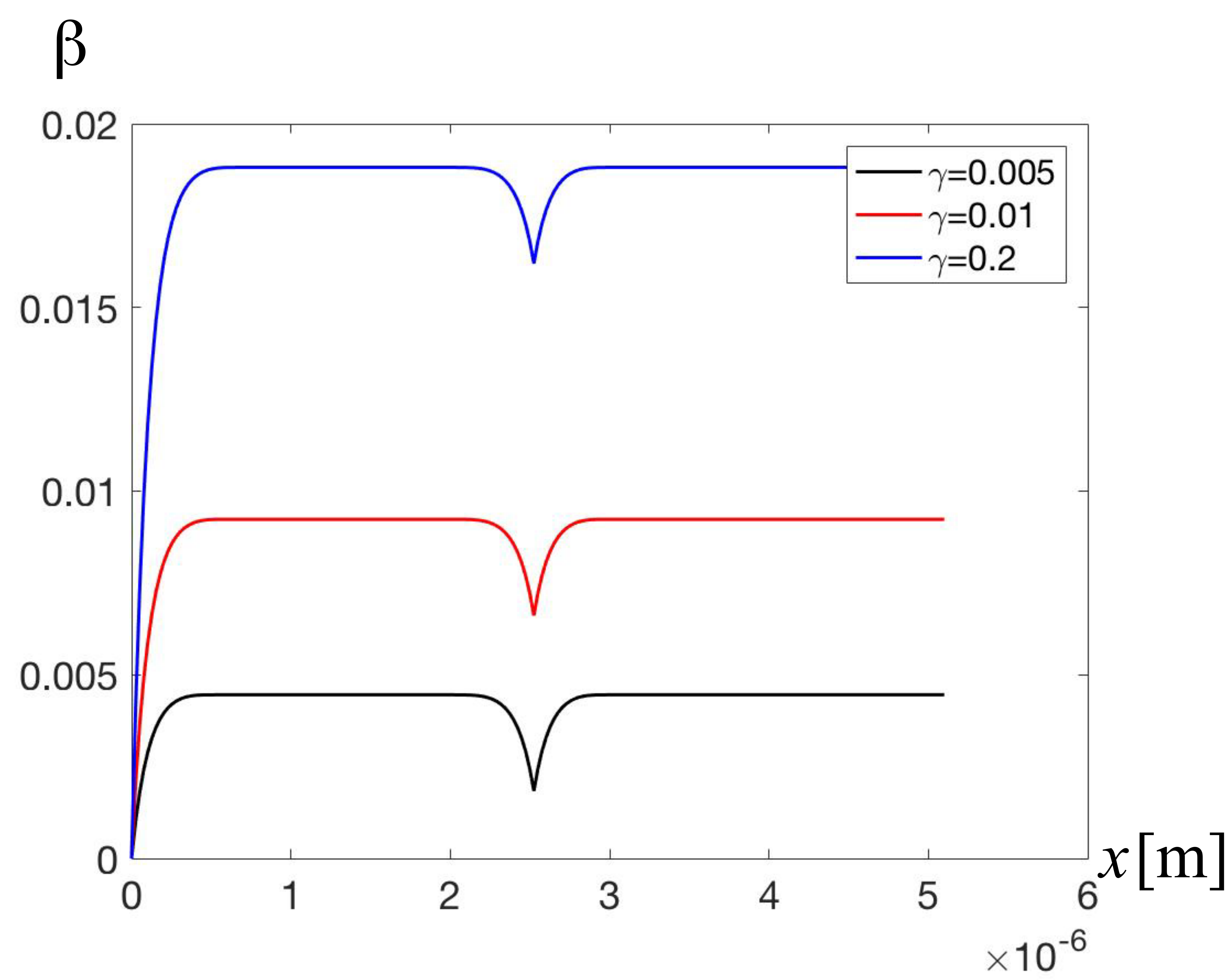} \\
		(a)  &   (b)  \\[6pt]
	\end{tabular}
	\caption{(Color online) Plastic slips $\beta(x)$ at different shear strains: (a) $\gamma=0.001$, $\gamma=0.002$, $\gamma=0.003$ in the dislocation pile-up process; (b) $\gamma=0.005$, $\gamma=0.01$, $\gamma=0.02$ in the dislocation traversal process.}
	\label{Plastic Slips}
\end{figure}

With this grain boundary condition, the simulated plastic slip distributions at three different shear strains in the pile-up process and in the traversal process are shown in Fig.~\ref{Plastic Slips}(a) and \ref{Plastic Slips}(b), respectively. The homogeneous Dirichlet boundary condition $\beta(0)=0$ causes the plastic slip in the left boundary layer to change rapidly, while the Neumann boundary condition $\beta_{,x}(c)=0$ keeps the plastic slip near the right free surface constant. At the position of the grain boundary $x=c/2$ there is a groove due to Eq.~\eqref{Constraint}$_1$. In the process of dislocation pile-up against the grain boundary, the groove sinks as the strain increases (Fig.~\ref{Plastic Slips}(a)), while it retains its shape when the dislocation passing process starts (Fig.~\ref{Plastic Slips}(b)). During the load reversal, one sees hills instead of grooves.  Note that depending on the types of dislocations and the slip systems activated, the distribution of plastic slip in two adjacent grains may have distinct highest magnitudes \citep{kochmann2008dislocation,kochmann2009plastic}.

\begin{figure}[htp]
	\begin{tabular}{cc}
		\includegraphics[width=0.48 \textwidth]{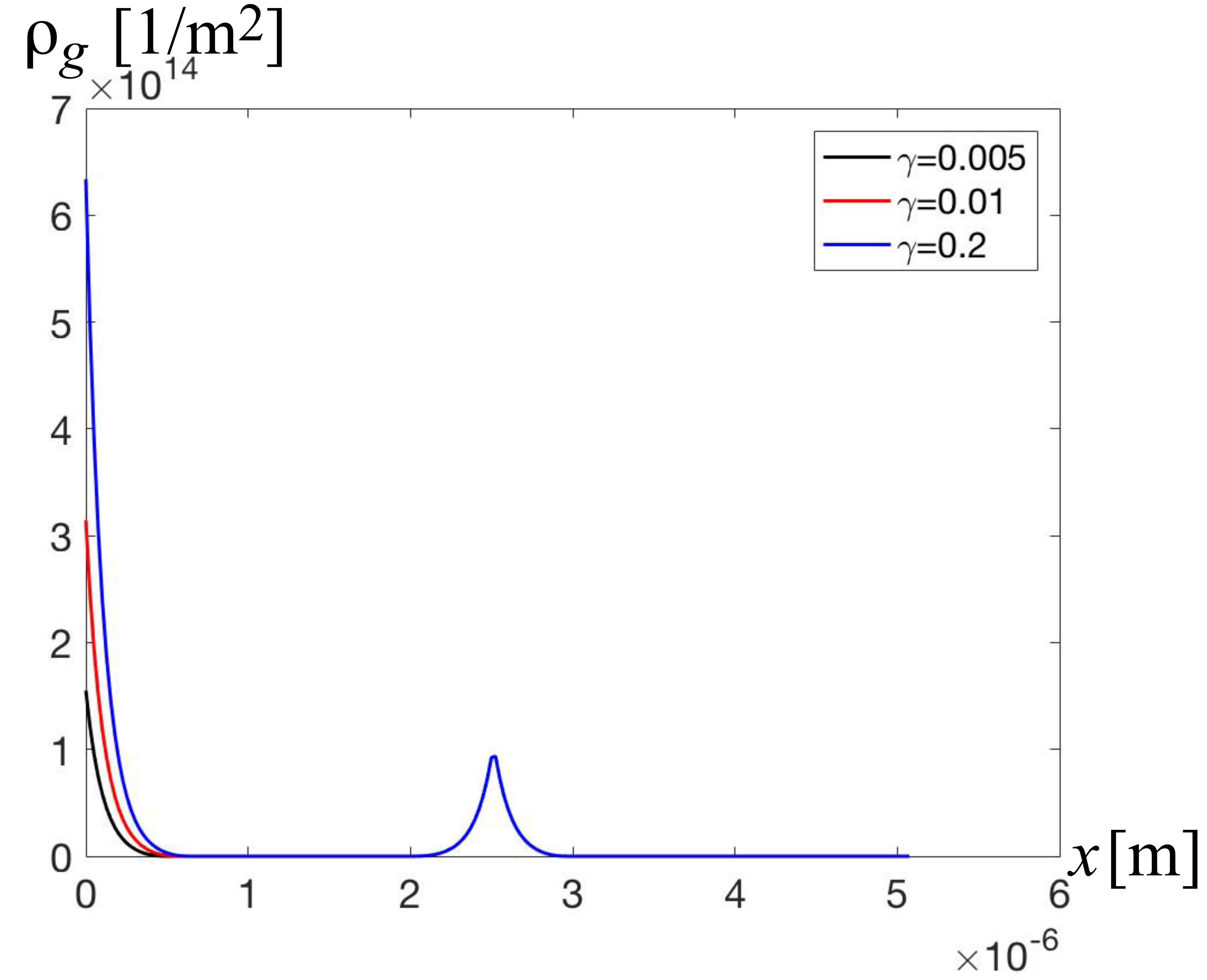} &  
		\includegraphics[width=0.45 \textwidth]{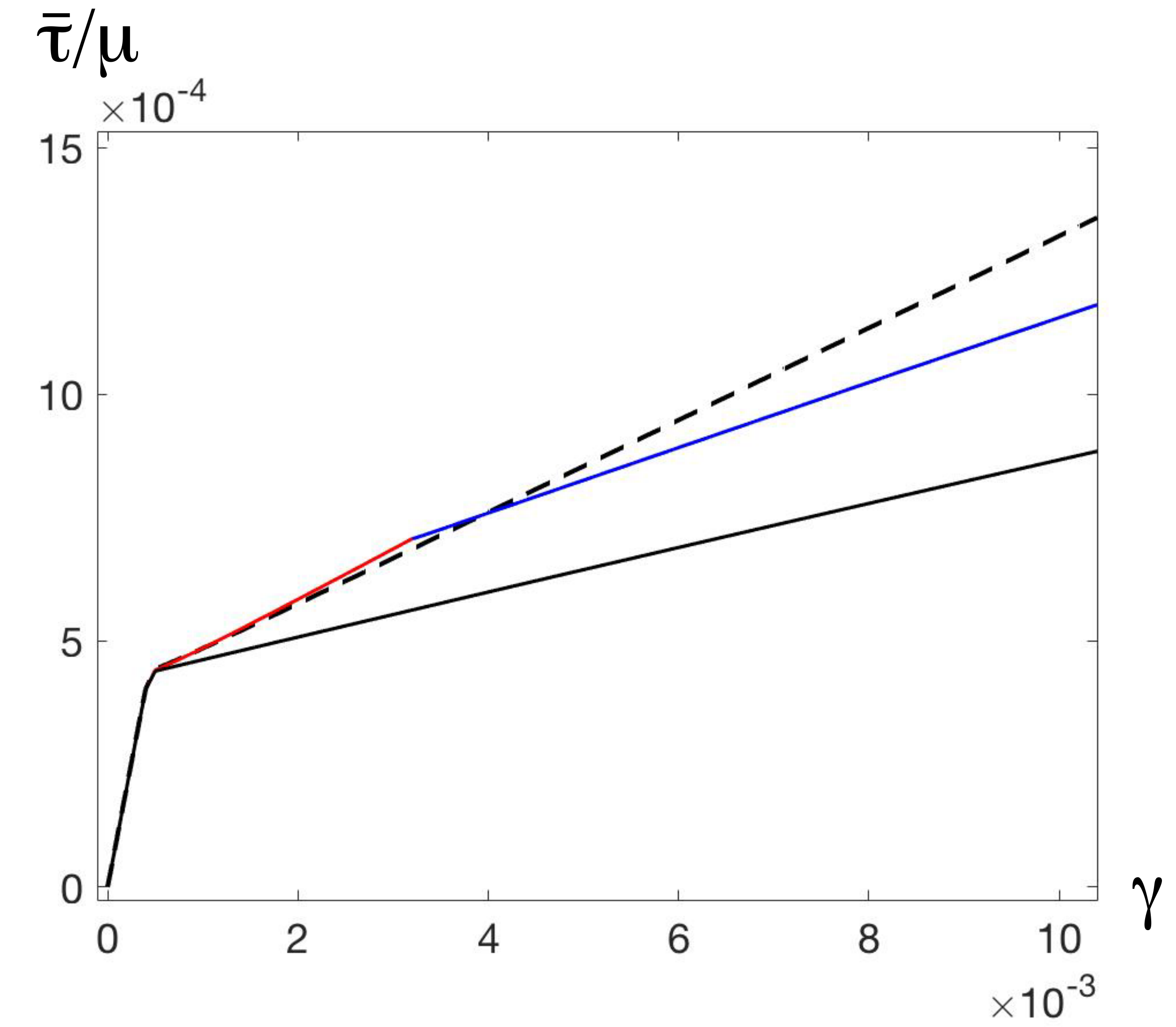} \\
		(a)  &   (b)  \\[6pt]
	\end{tabular}
	\caption{(Color online) (a) Distributions of non-redundant dislocation density $\rho_g$ at three different shear strains: $\gamma=0.005$, $\gamma=0.01$, $\gamma=0.02$. (b) Stress-strain curve: (i) Mixed hardening (colored), (ii) The hardening excluding the effect of back stress (bold black), (iii) The hardening reported in \cite{le2018cthermodynamic} (dashed black).  }
	\label{Hardening and DislocationDensity}
\end{figure}

Initially, the crystal sample has only redundant dislocations in the form of dipoles. When the internal stress exceeds the Taylor stress at the initial yielding, dislocation dipoles begin to dissolve into positive and negative dislocations. Then, under the applied shear stress, the positive dislocations move to the left and the negative dislocations to the right. In the $(0, c/2)$-interval, positive dislocations pile up against the left surface and negative dislocations against the left side of the grain boundary, while in the $(c/2, c)$-interval, the positive dislocations against the right side of the grain boundary and the negative dislocations leave the specimen through the right free surface. During the pile-up process the distribution of non-redundant dislocations in the left boundary layer and near the grain boundary is symmetrical but differs in the traversal process. In the latter process, as the applied strain increases, both the magnitude of the non-redundant dislocation density and the width of the dislocation-occupied zone in the left boundary layer increase but remain identical at the grain boundary because overflowed positive and negative non-redundant dislocations cancel each other out and become redundant (see Figure~\ref{Hardening and DislocationDensity}(a) where three curves coincide at the grain boundary).

The averaged normalized shear stress $\bar{\tau}/\mu=\frac{1}{\mu c}\int_0^c \tau (x)\dd{x}$ against the shear strain $\gamma$ curve (further called stress-strain curve for short) describing the work hardening under the boundary conditions \eqref{Constraint} is shown as a colored curve in Fig.~\ref{Hardening and DislocationDensity}(b). Note that this averaged shear stress is directly related to the load $F$ measured in experiments, as $F$ is nothing else but $\bar{\tau}$ times the area $cL$. For comparison, we plot the isotropic work hardening, which ignores the contribution of the back stress, by bold black and the work hardening of a sample exposed to two Dirichlet boundary conditions at two ends (\citep{le2018cthermodynamic}, without grain boundary), by dashed black. Looking at the colored curve, we see that the first stage, represented by a linear line with a high slope, is an elastic zone, and the rest is a plastic zone in which two hardening slopes are shown. The reason for two work hardening slopes is that in the second stage, marked by red, non-redundant dislocations pile up against both left surface and grain boundary, while in the third stage, marked by blue, the effect of hardening of non-redundant dislocations at the grain boundary has elapsed so that the work hardening slope is reduced. All three curves coincide in the elastic zone but diverge as the plastic deformation develops. The difference between the two upper curves and the bold black curve is the kinematic work hardening, which becomes remarkable at high strains. If one compares the colored and the dashed curve, one can see that the former becomes stronger in the second stage, but less hard in the third. This is due to the change in the number of locations where the dislocation pile-up occurs. In accordance with the dislocation impediment by the grain boundaries, there are three places in the second stage and one in the third stage for the colored curve, whereas there are always two in the dashed curve throughout the process.

As observed by \citet{kondo2016direct}, dislocations whose lines are almost parallel to the plane of the grain boundary move towards the low angle grain boundary and are slightly hindered. In a high angle grain boundary, the plane is inclined to the dislocation line and dislocations are not detected to cross the grain boundary until the edge of the specimen has been broken due to the high stress at the indentation point. It is reasonable to assume that there is a moderate angle grain boundary which can impede a higher dislocation density than a low angle grain boundary. This is equivalent to saying that $\zeta _Y$ in the surface dissipation has three values depending on whether the misorientation angle is small, moderate, or large. Eq.~\eqref{grain_bc} then yields two critical dislocation densities $\rho_{cr1}$ and $\rho_{cr2}$ for small and moderate misorientation angles, while it has no root for large misorientation angles. If we assume the critical dislocation density impeded by the mid-angle grain boundary to be $\rho_{cr2}=2 \rho_{cr1}$, then the conditions at this boundary are
\begin{equation}\label{Constraint2}
\begin{cases}
\dot{\beta}(c/2,t)=0 & \text{as long as $\rho_g < \rho_{cr2}$}, \\
|\beta_{,x}(c/2\pm 0,t)| =b\rho_{cr2} & \text{otherwise}.
\end{cases}
\end{equation}

\begin{figure}[htp]
	\begin{tabular}{cc}
		\includegraphics[width=0.48 \textwidth]{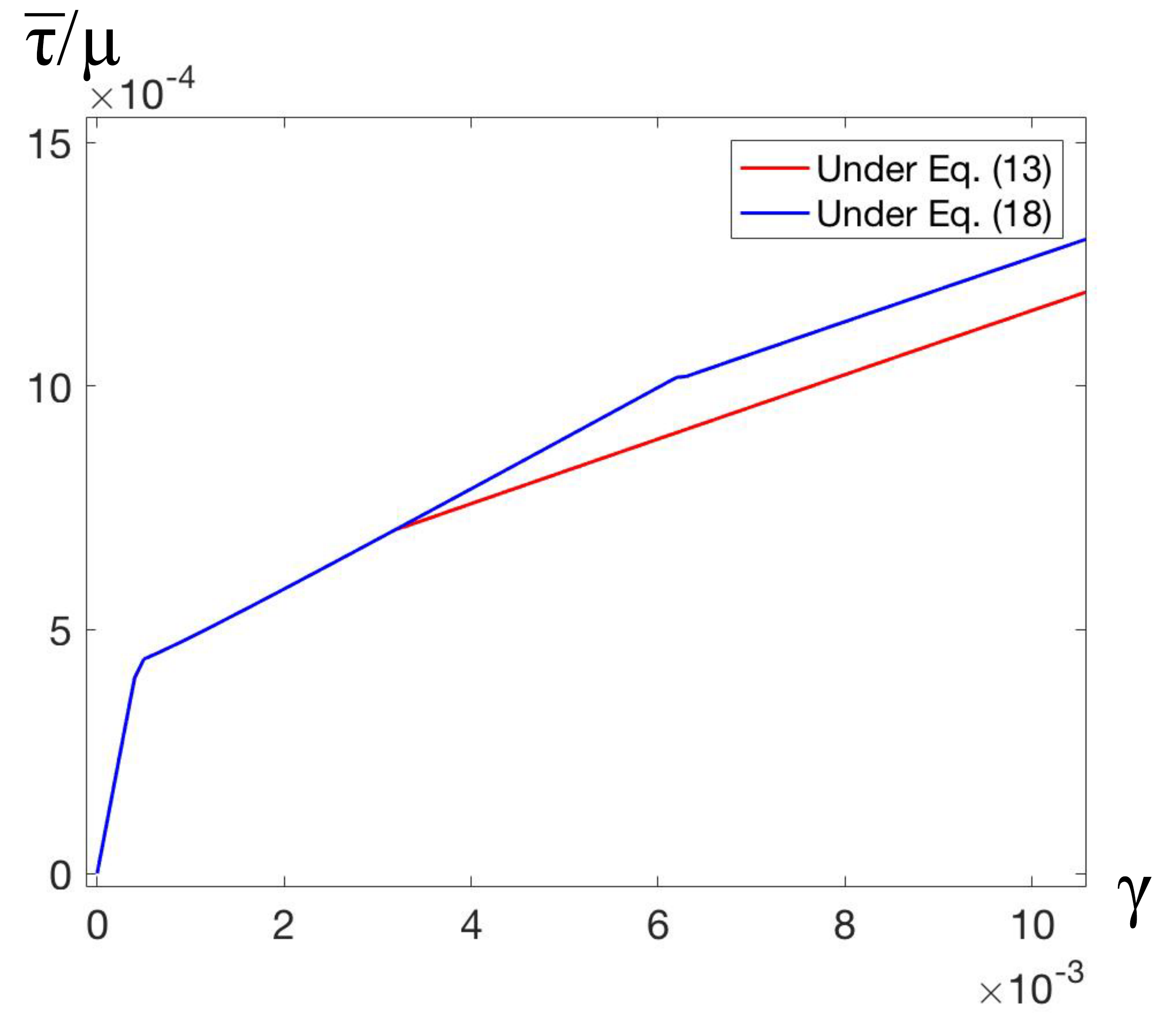} &  
		\includegraphics[width=0.48 \textwidth]{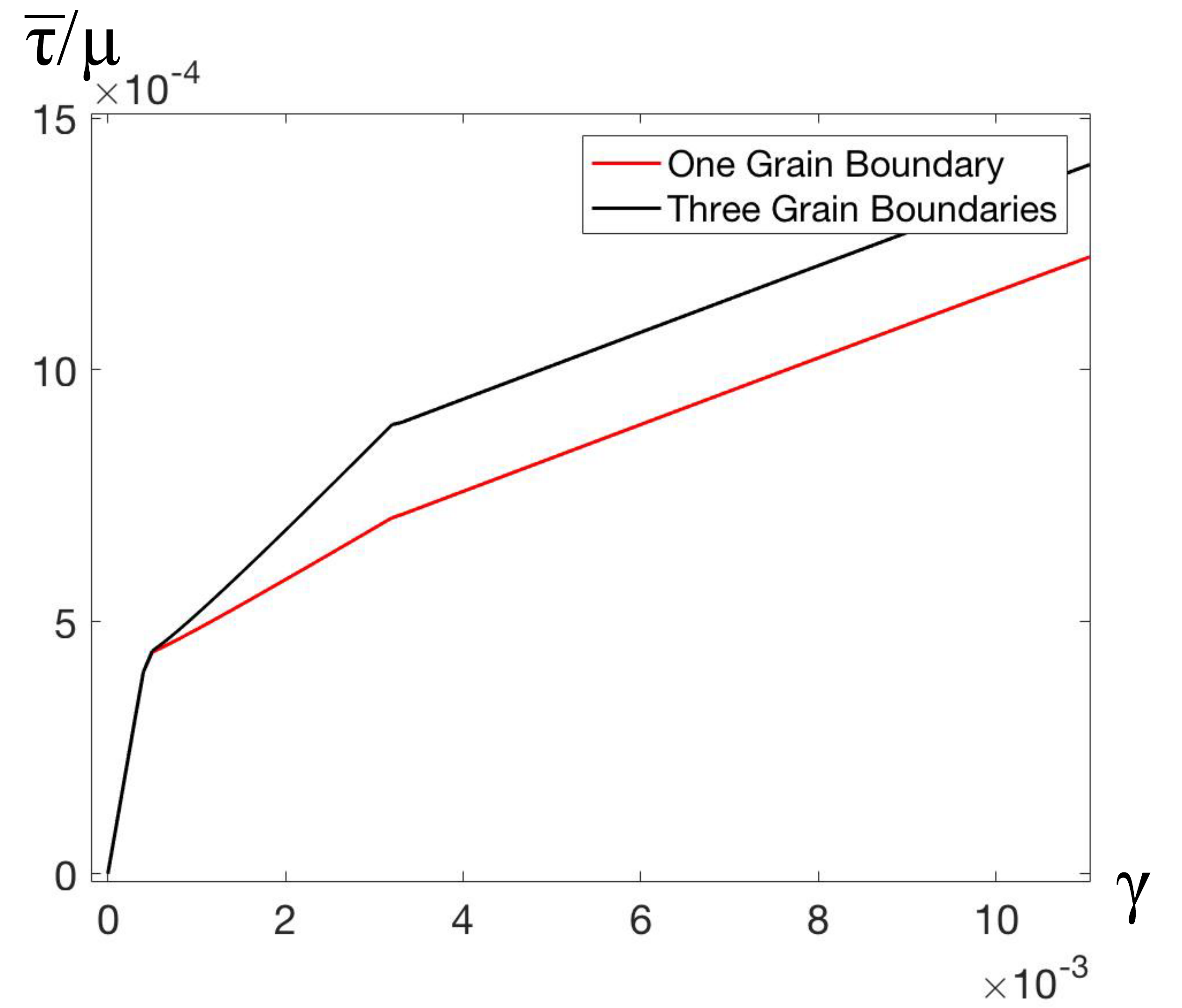} \\
		(a)  &   (b)  \\[6pt]
	\end{tabular}
	\caption{(Color online) Comparing two stress-strain curves: (a) Sample under Eq.~\eqref{Constraint} and sample under Eq.~\eqref{Constraint2}. (b) Sample with one grain boundary and sample with three grain boundaries. }
	\label{Two Influence}
\end{figure}

In Fig.~\ref{Two Influence}(a) the work hardening of two bi-crystals, one with a low-angle and one with a mid-angle grain boundary at $x=c/2$, are compared. The result shows that the hardening rate of each stage is the same for two samples, but the length of the second stage of the blue curve is longer than that of the red curve because the critical density in the traversal process is twice as high (see Eq.~\eqref{Constraint2}$_2$). The reason for this is that a higher elongation is required for the mid-angle grain boundary to reach its critical density. Next, we simulated the stress-strain curve of a polycrystal with three low-angle grain boundaries at $x=c/4$, $x=c/2$, $x=3c/4$ and compared it to the previous simulation of a sample with a low-angle grain boundary (Fig.~\ref{Two Influence}(b)). Since the same boundary condition \eqref{Constraint} is applied to all three grain boundaries, in this case the length of each step is the same for two samples, but the hardening rates of the second step are different. The more grain boundaries involved in the crystal, the higher the second stage hardening rate. The third stage hardening rates are identical for two specimens as they only depend on dislocations that pile up against the left surface.

\begin{figure}[htp]
	\begin{tabular}{cc}
		\includegraphics[width=0.44 \textwidth]{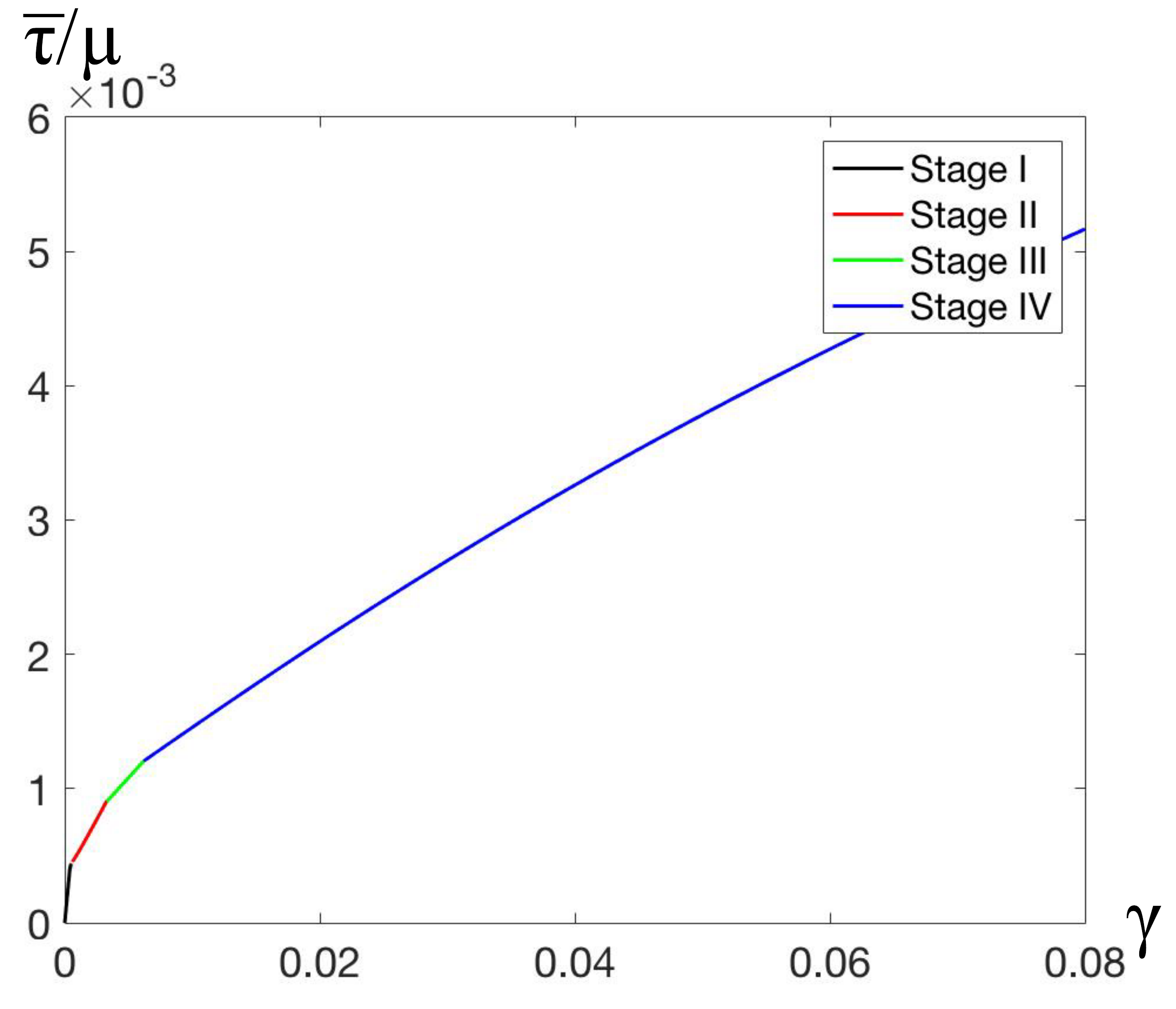} &  
		\includegraphics[width=0.48 \textwidth]{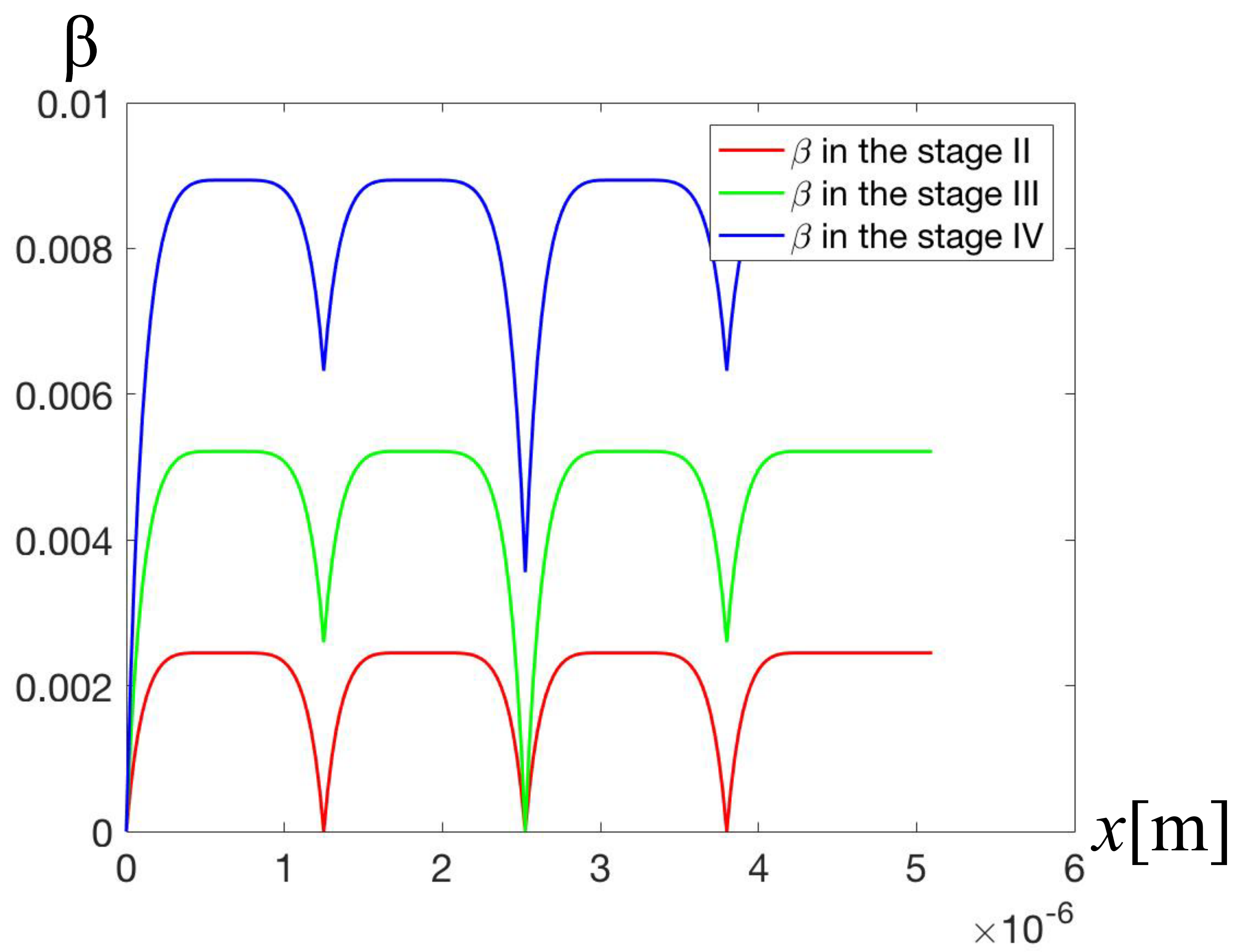} \\
		(a)  &   (b)  \\[6pt]
	\end{tabular}
	\caption{(Color online) (a) Stress-strain curve for a sample possessing two types of grain boundaries. (b) Distribution of plastic slip $\beta$ at each stage of plastic loading. }
	\label{HardeningMultiSlope}
\end{figure}

Fig.~\ref{HardeningMultiSlope}(a) shows the work hardening of a polycrystal that has two types of grain boundaries. Eqs.~\eqref{Constraint} are applied to two low-angle grain boundaries at $x=c/4$ and $x=3c/4$, while Eqs.~\eqref{Constraint2} are applied to a mid-angle grain boundary at $x=c/2$. In accordance with the dislocation impediment by the grain boundaries, after elastic loading, three boundaries and the left surface participate in the dislocation pile-up process in the second stage (red colored section). When the non-redundant dislocation density at the grain boundaries reaches $\rho_{cr1}$, the third stage begins (green colored section). In this stage, two low angle grain boundaries do not contribute to the increase in back stress but the left surface and moderate angle grain boundary do. Similarly, beyond $\rho_{cr2}$ only the left surface is responsible for kinematic hardening in the last stage (colored blue). The distributions of plastic slip for three stages during plastic deformation are shown in Fig.~\ref{HardeningMultiSlope}(b), where the grooves represent the dislocation pile-ups at the grain boundaries. If the groove is growing but its tip is still connected to the bottom, this means that non-redundant dislocations continue to accumulate at the grain boundary. When the groove tip leaves the bottom, this is related to the process of dislocation crossing at the corresponding grain boundary. Note that the hardening curve has more steps and the transition step is smoother at low strain when more conditions of a similar nature but of different magnitude are involved.

\begin{figure}[htp]
	\begin{tabular}{cc}
		\includegraphics[width=0.48 \textwidth]{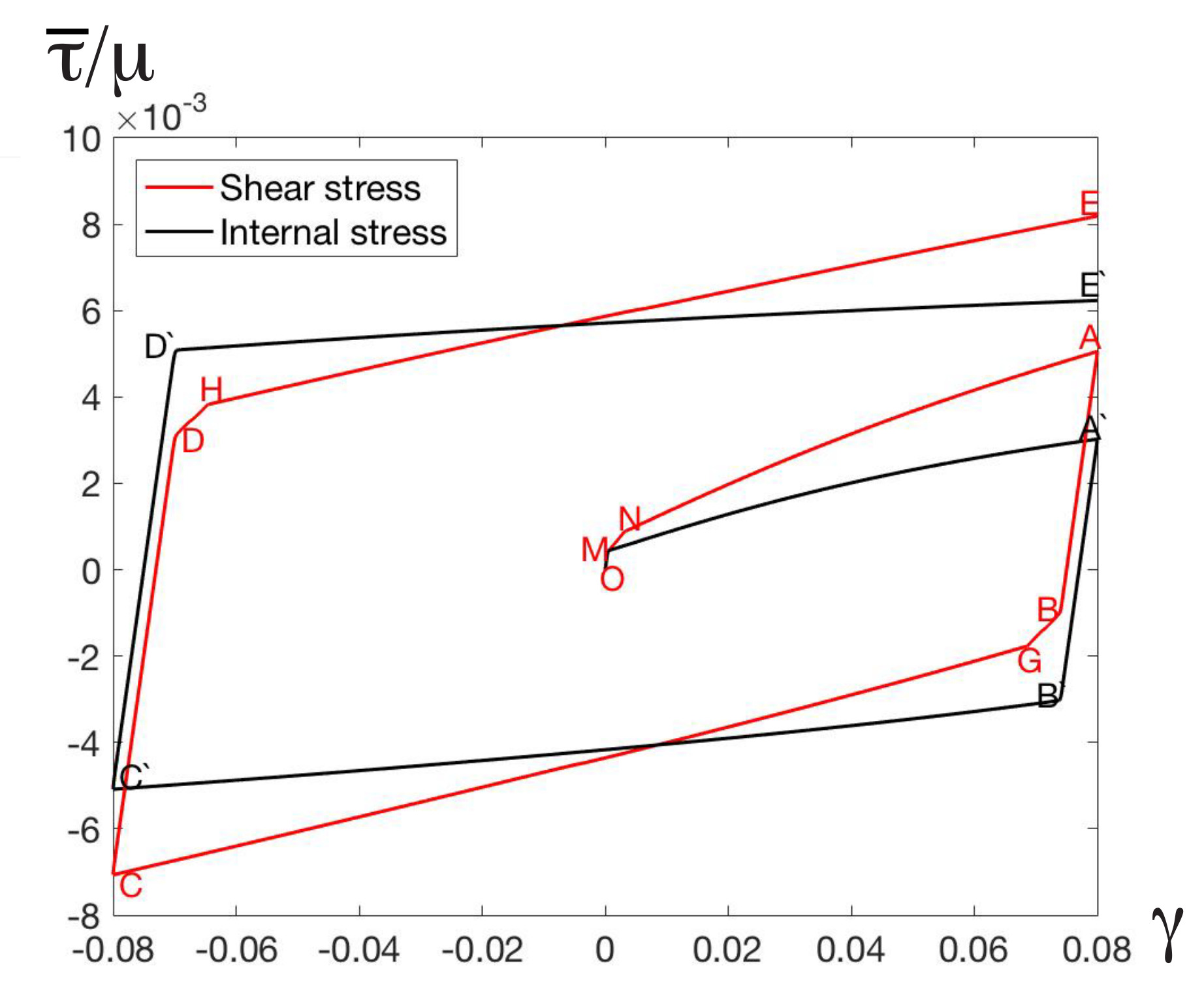} &  
		\includegraphics[width=0.46 \textwidth]{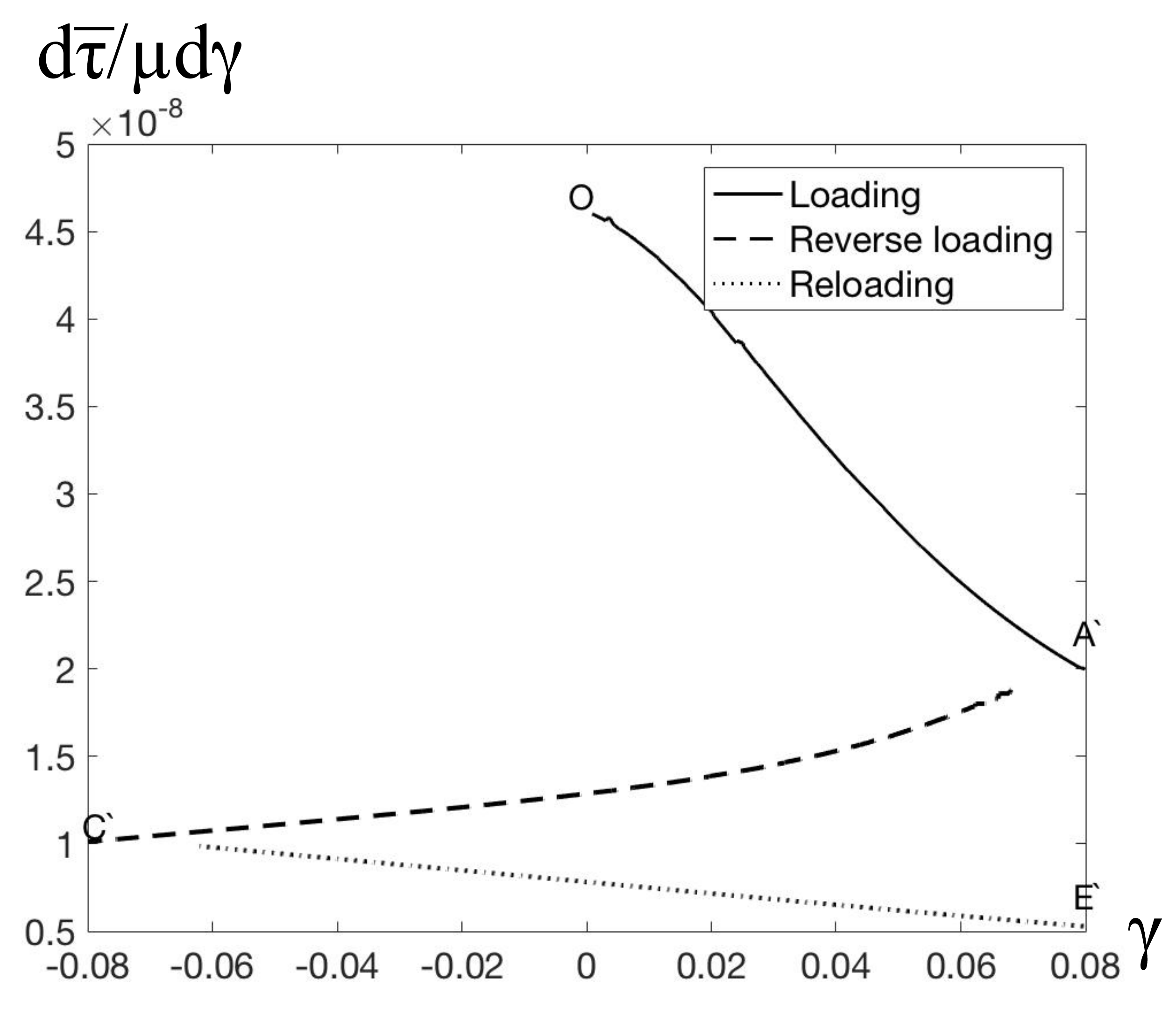} \\
		(a)  &  (b) \\[6pt]
	\end{tabular}
	\caption{(Color online) (a) Shear stress (red) and internal stress (black) versus shear strain for one and half cyclic loading. (b) The hardening rate as function of shear strain $\gamma$. }
	\label{One Cycle}
\end{figure}

Let us consider a load path that includes loading when $\gamma$ increases from 0 to $\gamma^*=0.08$, reverse loading from $\gamma^*$ to $-\gamma^*$, and reloading from $-\gamma^*$ to $\gamma^*$. This path is applied to a sample that has three low angle grain boundaries at $x=c/4$, $x=c/2$, $x=3c/4$. The corresponding cyclic stress-strain curve (red) is shown in Figure ~\ref{One Cycle}(a). Line AB and line CD are straight and parallel to each other as they correspond to the elastic load. This cyclic curve shows a Bauschinger effect as $|\tau_A| > |\tau_B|$ and $|\tau_C| > |\tau_D|$. \citet{le2018cthermodynamic} provided a physical explanation of the Bauschinger effect based on the back stress resulting from the energy density of non-redundant dislocations. Another interesting behavior of this curve is that the length of the transition states in the load reversal process (line BG) and in the re-loading process (line DH) are elongated compared to that in the loading process (line MN) due to the evolution of pile-ups of non-redundant dislocations at grain boundaries. This phenomenon can be observed in the experiment, which we will discuss in more detail in the next Section. The black cyclic curve is the stress-strain curve without the effect of back stress and shows isotropic work hardening behavior as $|\tau_{A'}| = |\tau_{B'}|$ and $|\tau_{C'}| = |\tau_{D'}|$. 

In both cyclic curves it can be observed that the work hardening decreases from process to process. The normalized hardening rate $\dv*{\bar{\tau}}{\gamma}/\mu$ corresponding to isotropic hardening is shown in Fig.~\ref{One Cycle}(b) where the initial part of each process is omitted because the corresponding high slope is due to the elastic strain and cannot be regarded as the work hardening. As shown, the curve decreases monotonically for the given load path. The reason for this is that the development of redundant dislocation density guarantees that it does decrease in all three processes, resulting in an increase in the dimensionless plastic slip rate $q(T,\tau_i,\rho_r)$. It is therefore necessary that the hardening rate is a monotonically decreasing function of $\gamma$ (see Eq.~\eqref{Governing tau_Y}). Note that the internal stress $\tau_i$, as the other argument of $q(T,\tau_i,\rho_r)$, rarely influences the dimensionless plastic slip rate compared to $\rho_r$, because $q(T,\tau_i,\rho_r)$ is a double exponential function of $\tau_i$. In contrast to isotropic hardening, kinematic hardening has no influence on the decrease in the work hardening tendency. Consequently, two claims can be made. One is that the slope of the hardening curve in the subsequent loading process cannot be higher than in the previous process. The second is that the higher the strain of a specimen, the lower the hardening rate in the reverse loading process. Both results can be confirmed by the experiment performed by \citet{thuillier2009comparison}. If the size of the specimen is small, down to a few microns or sub-microns where the back stress dominates, the decrease in work hardening tendency in the subsequent loading process is not pronounced unless the strain $\gamma^*$ becomes large.

\section{Comparison with experiments} \label{Section Comparison}
\subsection{Experiment}
\begin{figure}[htp]
		\centering
		\includegraphics[width=0.48 \textwidth]{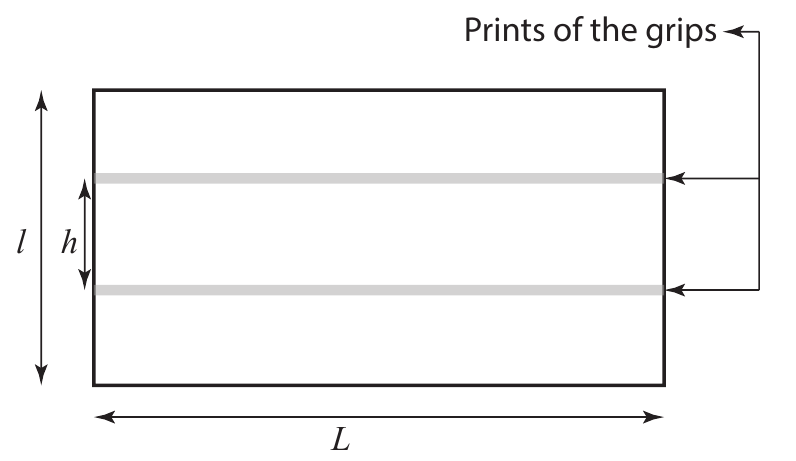} 
		\caption{Shear test sample.}
		\label{Shear device}
\end{figure}

\citet{thuillier2009comparison} performed shear tests on rectangular samples (Fig.~\ref{Shear device}) of dimension $L\times l= 50 \times 18$ mm$^2$ at the constant strain rate $\dot{\gamma}=2.1 \times 10^{-3} s^{-1}$. There is no precise information about the temperature, so we assume $T=298$ K. The shear strain $\gamma$ is measured as the maximum displacement in the shear direction divided by the gauge width, where the shear direction is along the length of the specimen. The samples have the shear gauge width $h$ of 4.5 mm and the sheet thickness $c_1$ of 0.7 mm. The studied material is a bake hardening mild steel E220BH whose Young's modulus and Poisson's ratio are $E=210$ GPa, $v=0.29$. The length of the sample is designed at least 10 times greater than the gauge width to prevent the edge effects. The sample is clamped between two grips, one attached to the fixed part and the other one to the moving part of apparatus. The sample is applied the reversed deformation until $\gamma=-0.4$ after different amounts of forward loading: $\gamma_1=0.1, \gamma_2=0.2, \gamma_3=0.3$. Each kind of test is performed three times to check the reproducibility.

\subsection{Parameter identification and numerical simulations}
In order to compute the theoretical stress-strain curves, we need values for several system-specific parameters. The basic parameters are the following: the activation temperature $T_P$, the stress ratio $r$, the steady-state scaled disorder temperature $\tilde{\chi}_0$, the two dimensionless conversion factors $K_{\rho}$ and $K_{\chi}$, the two coefficients $k_0$, and $k_1$ defining the function $\tilde{\tau}_b$. We also need initial values of the scaled dislocation density $\tilde{\rho}_i$ and the scaled disorder temperature $\tilde{\chi}_i$; both of which characterize the microstructure of the material prior to the plastic deformation and are determined by the sample preparation depending on many factors. The initial value of the dimensionless dislocation density determines the stress at which the onset of hardening occurs, and we use $\tilde{\rho}_i= 2.2\times 10^{-3}$ for a good fit to experimental results. The values of the coefficients $k_1$ and $k_0$ are given in \citep{le2018non} as $k_1=2.1\times 10^6, k_0=1 \times 10^{-6}$, and we set $\tilde{\chi}_0=0.25$ \citep{langer2010thermodynamic} and $\tilde{\tau}_i(0)=0$. The mean grain size $d$ determines the number of grain boundaries against which screw non-redundant dislocations pile up, and therefore, it is a crucial parameter in the simulation of the work hardening. Since the mean grain size of the sample was not measured in the experiment, an artificial value of $d=50 \, \mu$m is assigned.

\begin{figure}[htp]
	\centering
	\includegraphics[width=0.48 \textwidth]{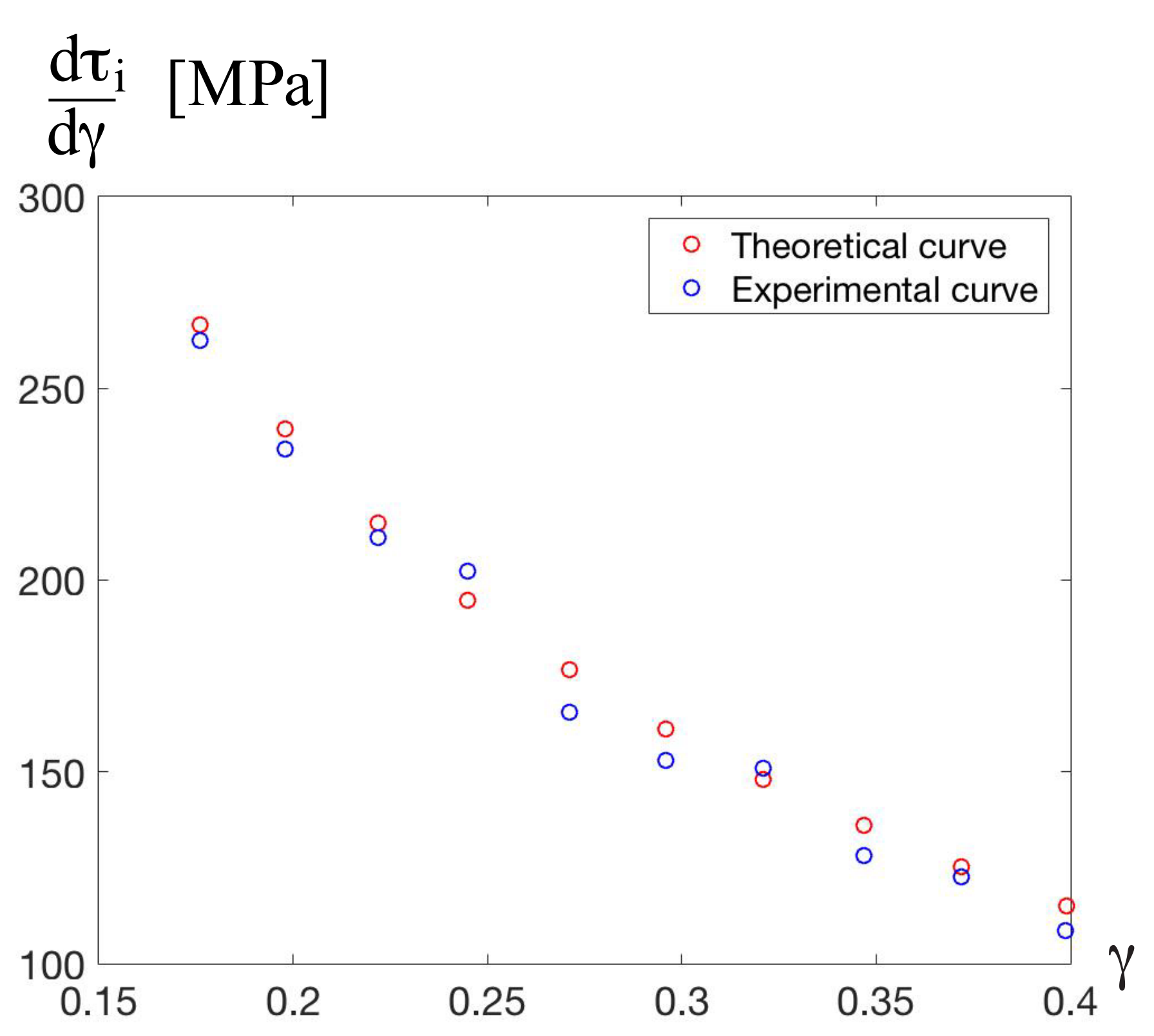}
	\caption{(Color online) Hardening rate of isotropic hardening in the interval $\gamma \in (0.18, 0.4)$. }
	\label{Slope of hardening}
\end{figure}

In earlier papers \citep{le2017thermodynamic,le2018bthermodynamic}, the desired system-specific parameters are evaluated by the large-scale least-squares analysis developed in \citep{le2018cthermodynamic}. That is, we have solved the discretized system of ordinary differential-algebraic equations (DAE) numerically, provided a set of the desired parameters and the initial values is known. Based on this numerical solution we then computed the sum of the squares of the differences between our theoretical stress-strain curves and a large set of selected experimental points, and minimized this sum in the space of the unknown parameters. However, the direct implementation of this method to the current investigation leads to an unmanageable computational cost because the condition subjected to the grain boundaries requires additional dozens of DAEs to be solved for running one iteration in the minimization process realized with the Matlab-globalsearch. 

\begin{figure}[htp]
	\centering
	\includegraphics[width=0.48 \textwidth]{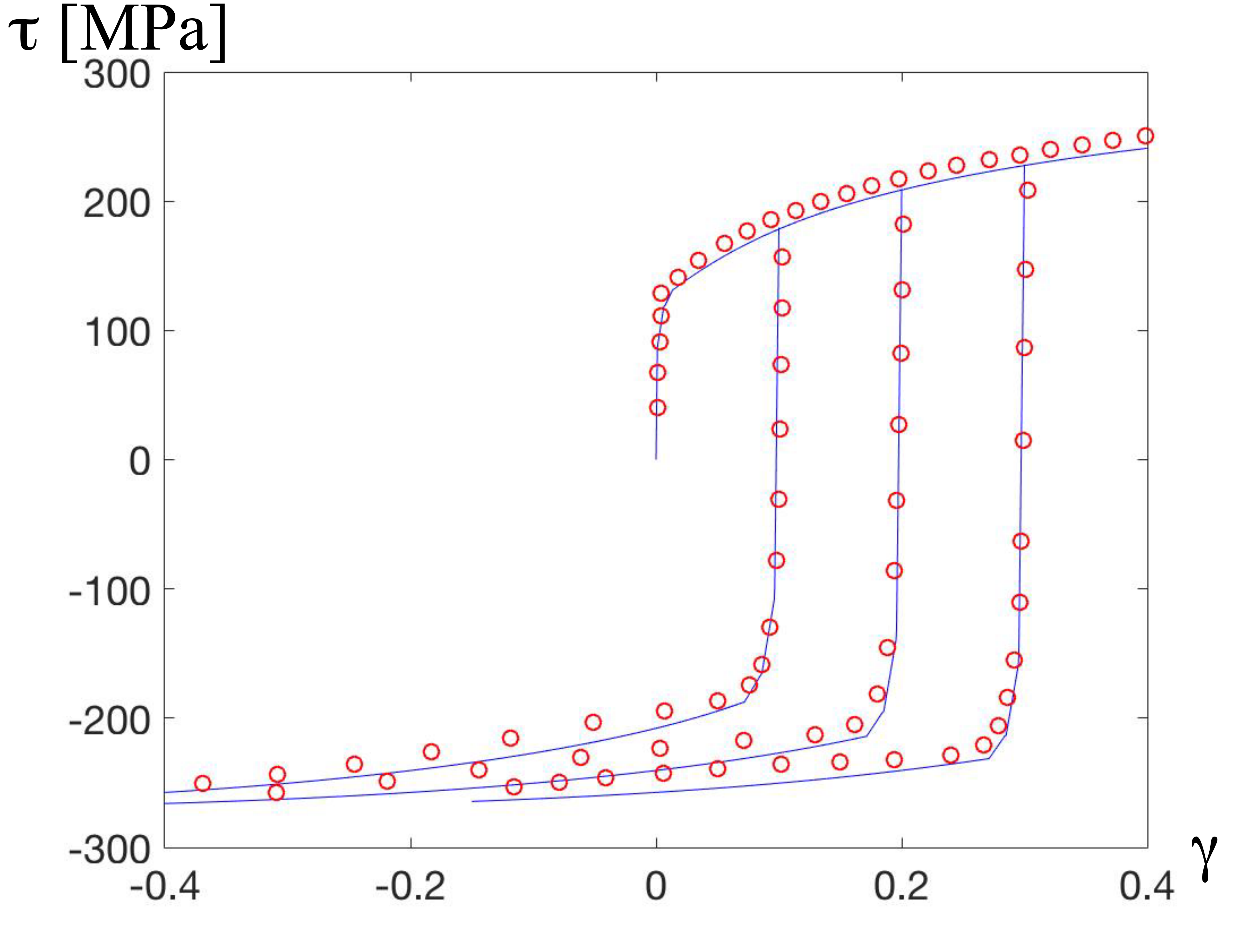} 
	\caption{(Color online) Stress-strain curve: (i) TDT (bold lines), (ii) experimental points taken from \citet{thuillier2009comparison} (circles). }
	\label{Stress-strain curve}
\end{figure}

Observing the stress-strain curves from the experiment (see Fig.~\ref{Stress-strain curve}), one can find that the stresses in three reversal processes approach a steady-state stress. The grain boundary prohibiting dislocation penetrating causes monotonically increasing kinematic hardening \citep{le2018non} both in the loading and reversal processes, while this phenomenon was not appeared in the experiments performed by \citet{thuillier2009comparison}. Therefore we assume that the specimens are free of the high angle grain boundaries, and since in such a case the dislocation impediment is only effective at the initial stage, the slope of the hardening curve at the later stage ($\gamma \in (0.18, 0.4)$) is only controlled by isotropic hardening. By comparing the slopes of the hardening for theoretical and experimental curves (Fig.~\ref{Slope of hardening}), the large-scale least-squares analysis becomes available to evaluate the remaining parameters because the computation associated with grain boundaries is excluded. The evaluated parameters are: $T_P=18024, r=0.0334, K_{\chi}=387.5, K_{\rho}=24.13, \tilde{\chi}_i=0.21$. Note that this strategy does not provide the unique set of the parameters for $T_P$ and $r$, namely, other combination of them may lead to the same slope curve but different magnitude of the stress-strain curve. For determining the magnitude, the average grain size and the corresponding kinematic hardening work as the complement.

The number of grain boundaries involved in the sample is $N_g=c_1/d-1$, and they are set to be homogeneously distributed along $x$-coordinate. Each grain boundary may possess different capacity of dislocation density, here we assume ten are low-angle grain boundaries with $\rho_{cr1}=0.57\times 10^{13}$ m$^{-2}$ and three mid-angle grain boundaries with $\rho_{cr2}=1.37\times 10^{13}$ m$^{-2}$. The magnitude of $\rho_{cr2}$ is verified according to the length of transition stage. We show our theoretical results based on the TDT together with the experimental data in Fig.~\ref{Stress-strain curve}. In this Figure, the circles represent the experimental data, while the bold lines are our theoretical simulation. It should be emphasized that the position of grain boundaries does not influence the stress-strain response. In other words, the size of grains have no impact on the kinematic hardening, but the number of grain boundaries plays a crucial role. This study tells us that the ``size'' in the size effect should be the average grain size that depends on the number of grains. This also means that the back stress produced by dislocation pile-up at the grain boundary is stronger for small size sample than for the large size sample, but it does not hold true for grains.

\begin{figure}[htp]
	\centering
	\includegraphics[width=0.48 \textwidth]{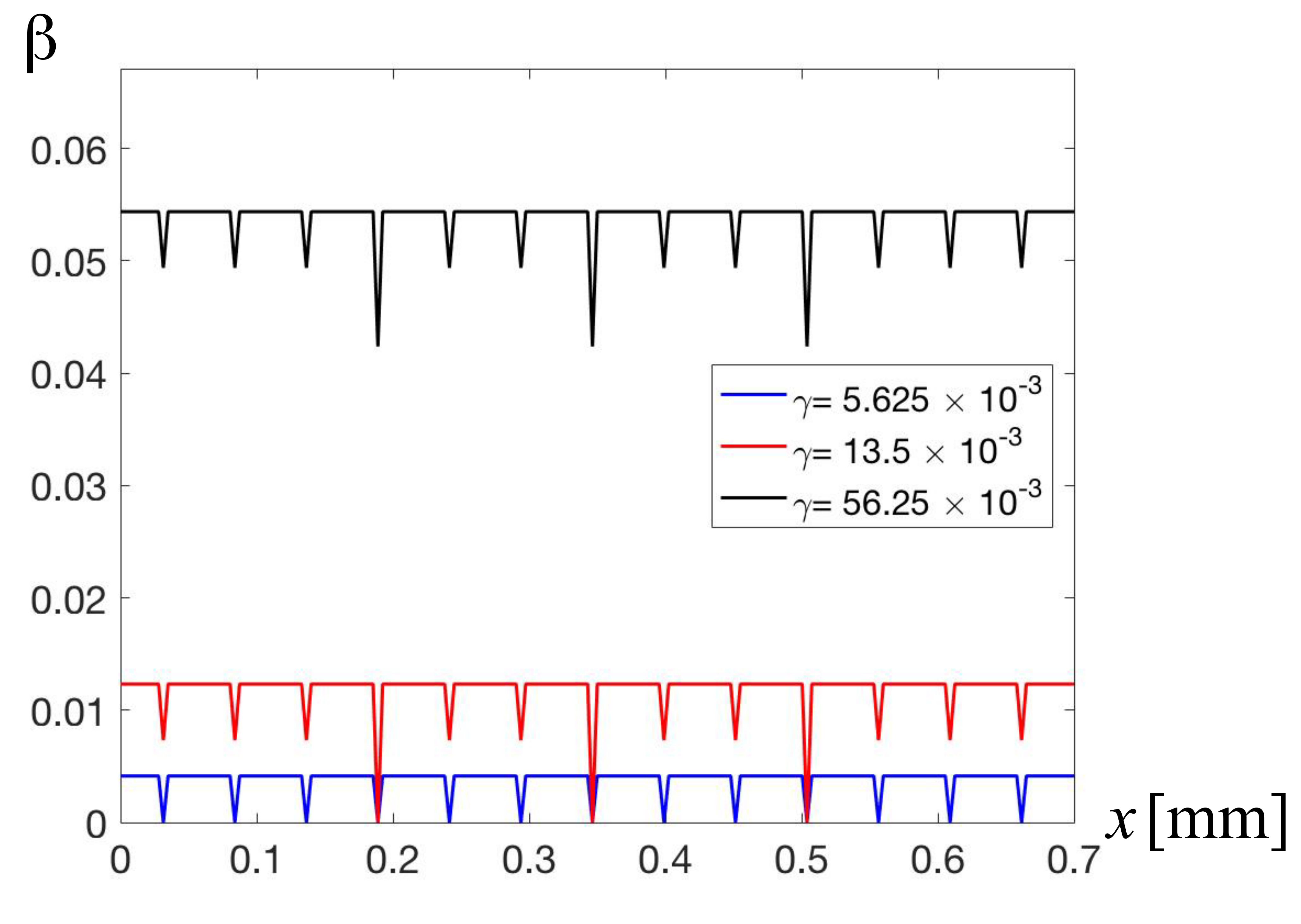} \\
	\caption{(Color online) Distributions of plastic slip in the loading process at $\gamma=5.625 \times 10^{-3}$, $\gamma=13.5\times 10^{-3}$, $\gamma=56.25\times 10^{-3}$. }
	\label{Plastic slips Loading}
\end{figure}

We plot in Fig.~\ref{Plastic slips Loading} three representative distributions of plastic slip at three different shear strains in the loading process. The evolution remains the same compared to that in Fig.~\ref{HardeningMultiSlope}(b), but the distribution is different due to the Neumann boundary condition $\beta_{,x}=0$, which is set at both the left and right free boundaries (to be consistent with the experimental setting) and the additional grain boundaries. Under the applied shear strain, positive non-redundant screw dislocations move towards left surface and the negative towards the right. 
\begin{figure}[t]
	\begin{tabular}{cc}
		\includegraphics[width=0.48 \textwidth]{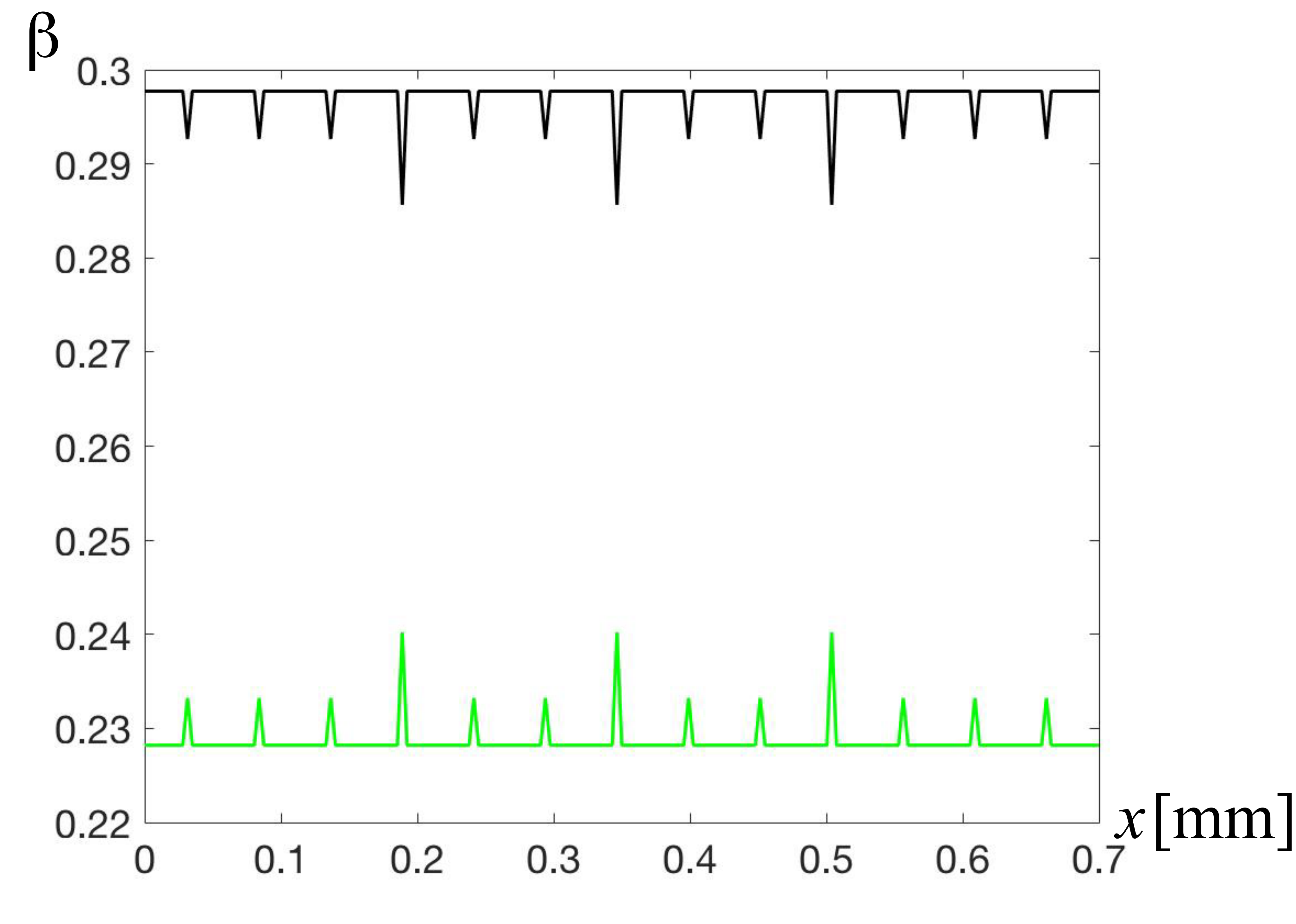} &
		\includegraphics[width=0.48 \textwidth]{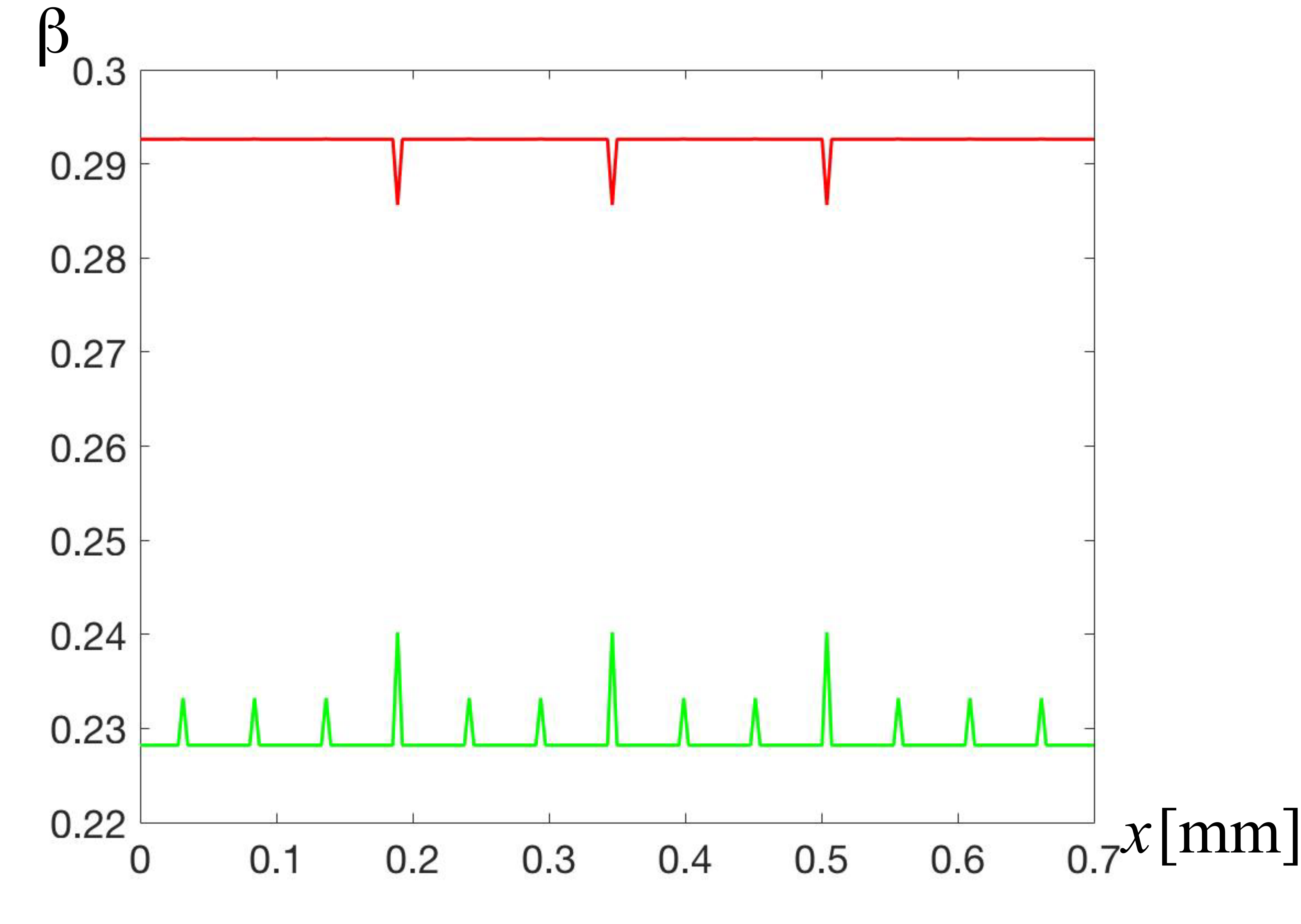} \\
		(a) & (b)\\
		\includegraphics[width=0.48 \textwidth]{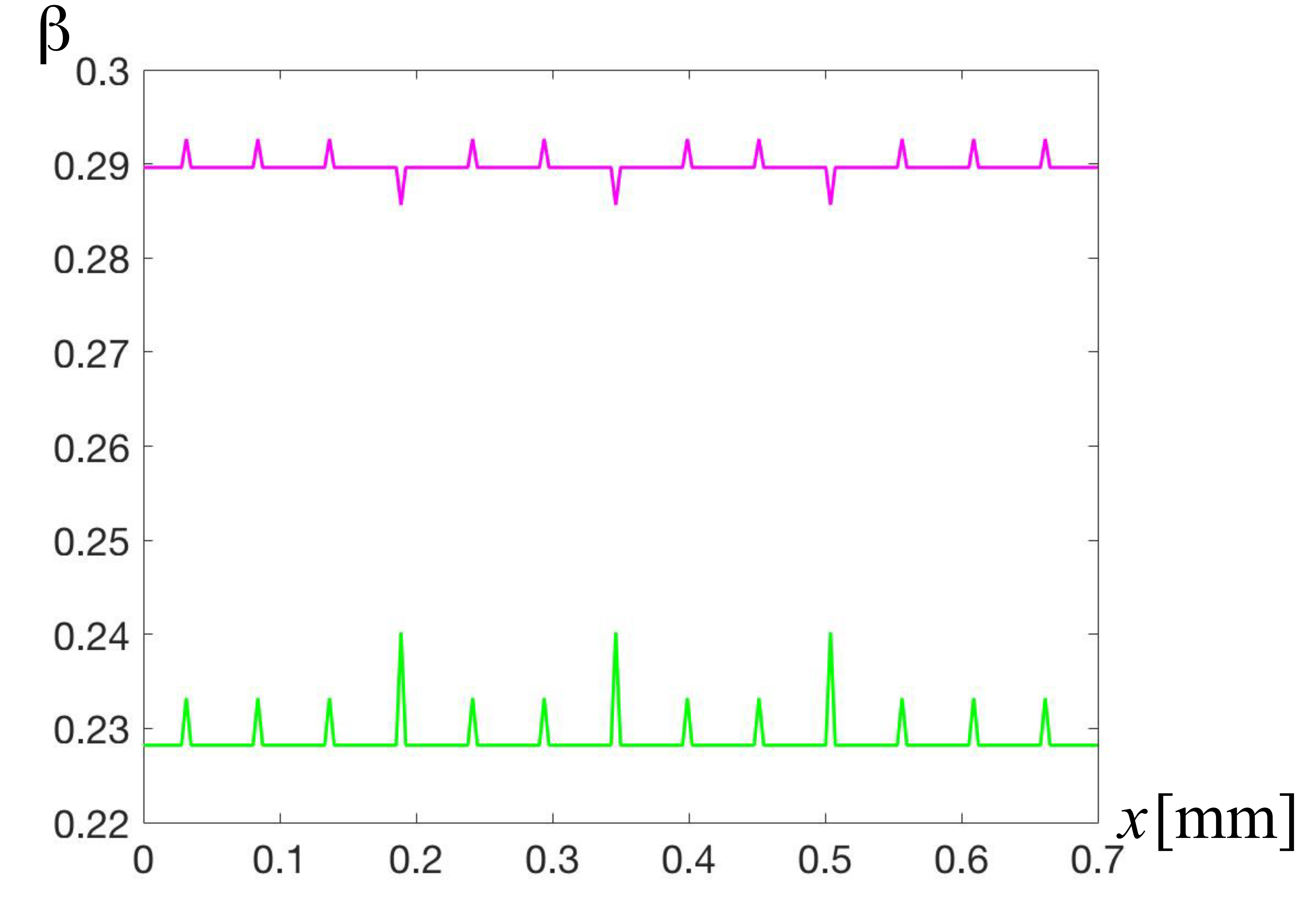} &
		\includegraphics[width=0.48 \textwidth]{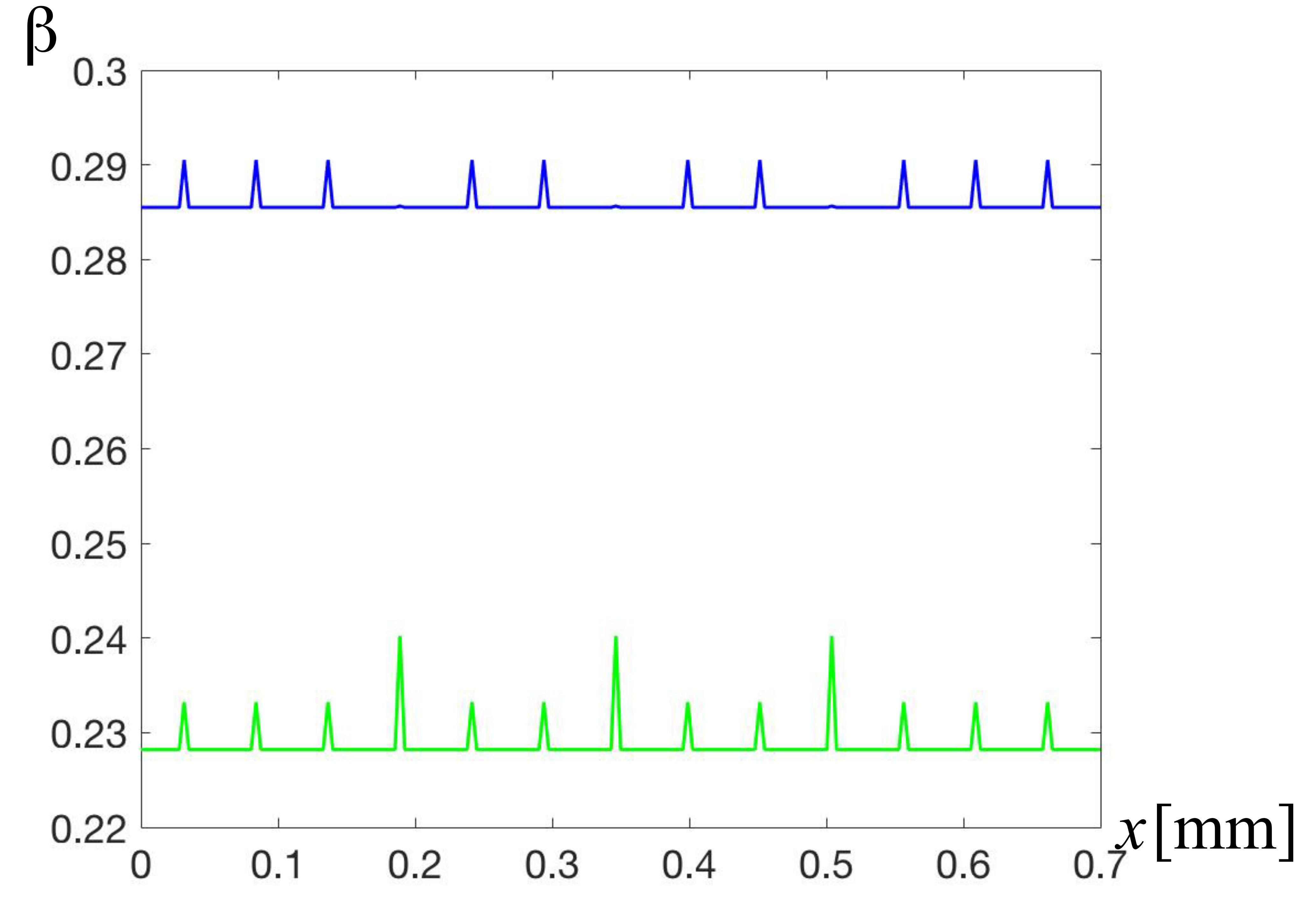} \\
		(c) & (d)
	\end{tabular}
	\caption{(Color online) Distributions of plastic slip in the reversal process: (a) $\gamma=\gamma_3$ (black), (b) $\gamma=\gamma_3-1.05\times10^{-2}$ (red), (c) $\gamma=\gamma_3-1.35\times10^{-2}$ (magenta), (d) $\gamma=\gamma_3-1.65\times10^{-2}$ (blue), and $\gamma=\gamma_3-7.5\times10^{-2}$ (green). }
	\label{Plastic slips Reversal}
\end{figure}

The distribution of plastic slip at $\gamma_3=0.3$ in the loading process is frozen during the elastic deformation of the reversal process and it is plotted in black in Fig.~\ref{Plastic slips Reversal}(a). Once the plastic deformation takes place, the plastic slip starts to evolve. In the following procedure, with the decreasing shear strain non-redundant dislocations move oppositely as the positive towards right and the negative left. As a result, the existing pile-up dislocations dismiss and the densities of non-redundant dislocations near the grain boundaries decrease. Non-redundant dislocations around low-angle grain boundaries come to a point that all are annihilated earlier than that happens near the moderate-angle grain boundary, and the plastic slip distribution has a shape of red curve in Fig.~\ref{Plastic slips Reversal}(b). Further reverse loading makes non-redundant dislocations of opposite sign accumulated at the position of low-angle grain boundaries, so that they form hills instead of grooves (magenta curve in Fig.~\ref{Plastic slips Reversal}(c)), and then come to the point that non-redundant dislocations are vanished at the mid-angle grain boundaries (blue curve, Fig.~\ref{Plastic slips Reversal}(d)). The green curve corresponding to the distribution of plastic slip after the transition stage, is plotted for comparison. The magnitude of this curve decreases as the shear strain in the opposite direction increases, but the shape retains. The evolution of plastic slip tells us that the first step after elastic deformation in the reversal process is associated with the disappearing of the existed non-redundant dislocations, such that the transition stage is elongated. The length of transition stage in the reversal process is about 2.3 times larger than that in the loading process. 

\begin{figure}[htp]
	\centering
	\includegraphics[width=0.48 \textwidth]{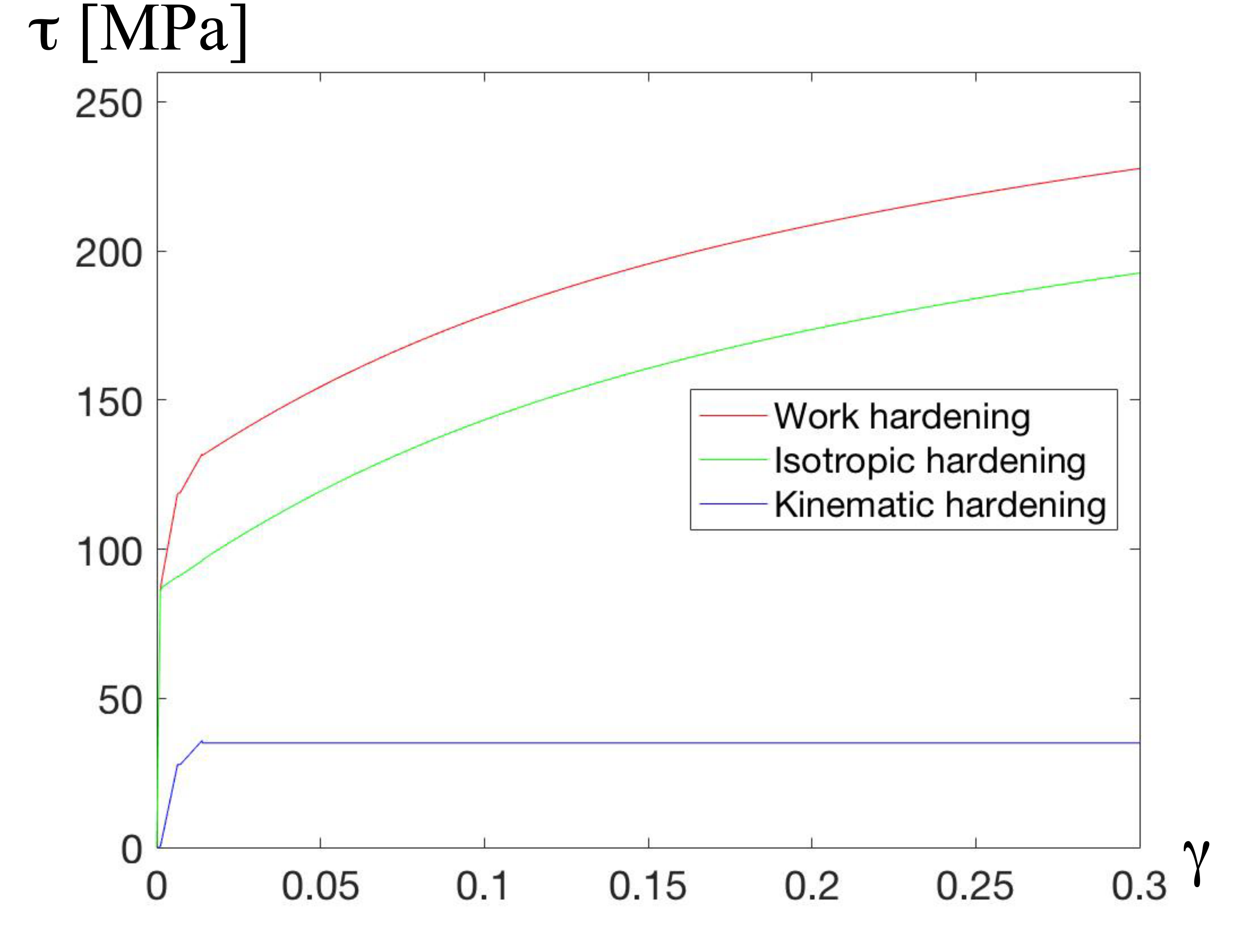} 
	\caption{(Color online) Partition of the work hardening.}
	\label{Partition Hardening}
\end{figure}

Fig.~\ref{Partition Hardening} presents the partition of the work hardening into the isotropic (green curve) and kinematic hardening (blue curve). The curve for the kinematic hardening shows that the magnitude of back stress increases rapidly for small strain and remain constant, which agrees with the observation by \citet{thuillier2009comparison} (blue dotted curve in Fig. 13 of their paper). For samples of smaller size and those possessing more grain boundaries, it could be expected that the contribution of kinematic hardening increases. 

\section{Conclusions and outlook} \label{Section Conclusion}
In this study, we found that the dimensionless plastic slip rate determined by Eq.~\eqref{TDT Dimensionless strain rate} describes correctly the behavior of isotropic work hardening, which decreases during plastic deformation and the load reversal process. For polycrystals with low and moderate angle grain boundaries, the average grain size rather than the size of individual grains affects the kinematic work hardening caused by the dislocation impediment. It is found that the number of grain boundaries determines the hardening rate and the ability of the grain boundaries to accumulate dislocations controls the length of work hardening at each stage. The annihilation of the existing non-redundant dislocations that accumulate at grain boundaries in the loading process leads to an extension of the transition stage in the load reversal process. 

We want to extend this research in at least two directions. First, we are going to apply the developed theory to the problem of twisted wires under load reversal with the aim of comparing the resulting Bauschinger effect with the experimental results reported in \citep{liu2013anomalous}. Second, we want to analyze the impediment of edge dislocations by the grain boundaries in polycrystals within the problem of plane constrained shear. When edge dislocations occurring in adjacent grains slide along different slip directions and pile up against the grain boundary, they cannot cancel each other out after reaching the grain boundary. In this case they form an array of superdislocations along the grain boundary which increase the misorientation angle and the jump in plastic slip. When this misorientation angle reaches a critical value, dislocations of another slip system occur in the adjacent grain, leading to the cross slip and the latent hardening. The major challenging block to the theory is that the cross slip can significantly affect the kinetics of dislocation depinning. These problems will be addressed in our forthcoming papers.

\section{Appendix} \label{Appendix}
In this Appendix the derivation of the governing equations of TDT from the variational equation \eqref{Sedov_equation} is provided. First, we compute the variation of the energy functional. Keeping in mind that $\rho_g=|\beta_{,x}|/b$ and $\rho=\rho_r+|\beta_{,x}|/b$, we have
\begin{multline}
\label{eq:A1}
\delta I=\int_{0}^c \Bigl[ -\mu (\gamma -\beta)\delta \beta+\gamma_D\delta \rho_r+\pdv{\psi_m}{\rho_g}\frac{1}{b}\text{sign}(\beta_{,x})\delta \beta_{,x}
\\
-(-\rho \ln (a^2\rho )+\rho )\delta \bar{\chi }
+\bar{\chi }\ln (a^2\rho )\Bigl( \delta \rho_r+\frac{1}{b}\text{sign}(\beta_{,x}) \delta \beta_{,x}\Bigr) \Bigr] \dd{x}.
\end{multline}
Decomposing this integral into two between $(0,c/2)$ and $(c/2,c)$ and integrating terms containing $\delta \beta_{,x}$ in \eqref{eq:A1} by parts, we obtain
\begin{multline}
\label{eq:A2}
\delta I=\int_{0}^c \Bigl[ -\mu (\gamma -\beta)\delta \beta+\gamma_D\delta \rho_r-\Bigl( \pdv{\psi_m}{\rho_g}\Bigr)_{,x} \frac{1}{b}\text{sign}(\beta_{,x})\delta \beta
\\
-(-\rho \ln (a^2\rho )+\rho )\delta \bar{\chi }
+\bar{\chi }\ln (a^2\rho )\delta \rho_r+\Bigl(\bar{\chi }\ln (a^2\rho )\Bigr)_{,x}\frac{1}{b}\text{sign}(\beta_{,x}) \delta \beta \Bigr] \dd{x}
\\
+\frac{1}{b}\Bigl( \pdv{\psi_m}{\rho_g}\text{sign}\beta_{,x}\Bigr|_{c/2-0} -\pdv{\psi_m}{\rho_g}\text{sign}\beta_{,x}\Bigr|_{c/2+0} +\bar{\chi }\ln (a^2\rho )\text{sign}\beta_{,x}\Bigr|_{c/2-0} 
\\
-\bar{\chi }\ln (a^2\rho ) \text{sign}\beta_{,x}\Bigr|_{c/2+0} \Bigr) \delta \beta(c/2)
+\frac{1}{b}\Bigl(\pdv{\psi_m}{\rho_g}+\bar{\chi }\ln (a^2\rho )\Bigr) \text{sign}\beta_{,x}\Bigr|_{c} \delta \beta (c).
\end{multline}
On the other side, the second integral in \eqref{Sedov_equation} equals
\begin{multline*}
\int_{0}^c \Bigl( \frac{\partial D_b}{\partial \dot{\beta}}\delta \beta +\frac{\partial D_b}{\partial \dot{\rho}}\delta \rho + \frac{\partial D_b}{\partial \dot{\bar{\chi}}}\delta \bar{\chi} \Bigr) \dd{x}
\\
=\int_{0}^c \Bigl[ \tau_i \delta \beta+d_\rho \dot{\rho} \Bigl( \delta \rho_r+\frac{1}{b}\text{sign}(\beta_{,x}) \delta \beta_{,x}\Bigr) +d_\chi \dot{\bar{\chi}} \delta \bar{\chi} \Bigr] \dd{x}.
\end{multline*}
Integration by parts of the term containing $\delta \beta_{,x}$ yields
\begin{multline}
\label{eq:A4}
\int_{0}^c \Bigl( \frac{\partial D_b}{\partial \dot{\beta}}\delta \beta +\frac{\partial D_b}{\partial \dot{\rho}}\delta \rho + \frac{\partial D_b}{\partial \dot{\chi}}\delta \chi \Bigr) \dd{x}=\int_{0}^c \Bigl[ \tau_i\delta \beta+d_\rho \dot{\rho} \delta \rho_r-\frac{1}{b}(d_\rho \dot{\rho})_{,x}\text{sign}(\beta_{,x})\delta \beta 
\\
+d_\chi \dot{\bar{\chi}}\delta \bar{\chi} \Bigr] \dd{x} +\frac{1}{b} \Bigl( d_\rho \dot{\rho}\text{sign}\beta_{,x}\Bigr|_{c/2-0} 
-d_\rho \dot{\rho}\text{sign}\beta_{,x}\Bigr|_{c/2+0} \Bigr) \delta \beta(c/2)+\frac{1}{b}d_\rho \dot{\rho}\text{sign}\beta_{,x}|_{c} \delta \beta (c)
\end{multline}
Substituting equations \eqref{eq:A2} and \eqref{eq:A4} into \eqref{Sedov_equation} and requiring that it is fulfilled for arbitrarily chosen $\beta$, $\rho_r$, and $\bar{\chi}$, we get
\begin{equation}
\label{eq:A5}
\begin{split}
-\tau -\Bigl( \pdv{\psi_m}{\rho_g}\Bigr)_{,x} \frac{1}{b}\text{sign}(\beta_{,x})- \frac{1}{b}(d_\rho \dot{\rho}+\bar{\chi }\ln (a^2\rho ))_{,x} \text{sign} \beta_{,x}+\tau _i=0, 
\\
\gamma_D+\bar{\chi }\ln (a^2\rho) +d_\rho \dot{\rho }=0, 
\\
-(-\rho \ln (a^2\rho) +\rho)+d_\chi \dot{\bar{\chi}}=0,
\end{split}
\end{equation}
where $\tau=\mu(\gamma-\beta)$ is the applied shear stress. By virtue of the second equation of \eqref{eq:A5} and the constancy of $\gamma_D$, the third term of the first equation vanishes. Introducing 
\begin{equation}
\label{eq:A6}
\tau_b=-\Bigl( \pdv{\psi_m}{\rho_g}\Bigr)_{,x} \frac{1}{b}\text{sign}(\beta_{,x})=-\pdv[2]{\psi_m}{\rho_g}\frac{1}{b^2}\beta_{,xx}
\end{equation}
as the back stress, we reduce the first equation of \eqref{eq:A5} to \eqref{TDT Balance of microforce}. The two remaining equations can be transformed to \eqref{Governing chi} and \eqref{Governing rho} if we choose
\begin{equation}
\label{eq:A7}
\begin{split}
d_\chi =\frac{\rho -\rho \ln (a^2\rho) }{LK_{\chi}\frac{\tau_i  e_D  q}{\mu  t_0}\Bigl[1-\frac{\chi}{\chi_0}\Bigr]},
\\
d_\rho =\frac{-e_D-\chi \ln (a^2\rho)}{LK_{\rho} \frac{\tau_i  q}{a^2 \mu \nu^2 t_0} \Bigl(1-\frac{\rho}{\rho_{ss}(\chi)}\Bigr)} . 
\end{split}
\end{equation}
Note that, for $\rho $ changing between 0 and $\rho _{ss}=\exp (-e_D/\chi )/a^2$, both numerators on the right-hand sides of \eqref{eq:A7} are positive, and the dissipative potential \eqref{TDT Dissipation function} is positive definite as required by the second law of thermodynamics. Taking Eqs.~\eqref{eq:A5} into account, we reduce the variational equation \eqref{Sedov_equation} to \eqref{reduced_variation}.


%
\section*{Conflict of interest}
The authors declare that they have no conflict of interest.



\end{document}